\documentclass[twocolumn]{aastex61}

\received{}
\revised{}
\accepted{}
\submitjournal{ApJ}

\shorttitle{Subaru HSC-FIRST Radio AGNs}
\shortauthors{Yamashita et al.}

\begin{document}

\title{
A Wide and Deep Exploration of Radio Galaxies with Subaru HSC (WERGS).\\
 I. the Optical Counterparts of FIRST Radio Sources}

\email{takuji@cosmos.phys.sci.ehime-u.ac.jp}

\author[0000-0002-4999-9965]{Takuji Yamashita}
\affil{Research Center for Space and Cosmic Evolution, Ehime University, 2-5 Bunkyo-cho, Matsuyama, Ehime 790-8577, Japan}

\author{Tohru Nagao}
\affiliation{Research Center for Space and Cosmic Evolution, Ehime University, 2-5 Bunkyo-cho, Matsuyama, Ehime 790-8577, Japan}

\author{Masayuki Akiyama}
\affiliation{Astronomical Institute, Tohoku University, 6-3 Aramaki, Aoba-ku, Sendai, Miyagi 980-8578, Japan}

\author{Wanqiu He}
\affiliation{Astronomical Institute, Tohoku University, 6-3 Aramaki, Aoba-ku, Sendai, Miyagi 980-8578, Japan}

\author{Hiroyuki Ikeda}
\affiliation{National Astronomical Observatory of Japan, 2-21-1 Osawa, Mitaka, Tokyo 181-8588, Japan}

\author{Masayuki Tanaka}
\affiliation{National Astronomical Observatory of Japan, 2-21-1 Osawa, Mitaka, Tokyo 181-8588, Japan}

\author{Mana Niida}
\affil{Graduate School of Science and Engineering, Ehime University, Bunkyo-cho, Matsuyama, Ehime 790-8577, Japan}

\author{Masaru Kajisawa}
\affiliation{Research Center for Space and Cosmic Evolution, Ehime University, 2-5 Bunkyo-cho, Matsuyama, Ehime 790-8577, Japan}
\affil{Graduate School of Science and Engineering, Ehime University, Bunkyo-cho, Matsuyama, Ehime 790-8577, Japan}

\author{Yoshiki Matsuoka}
\affil{Research Center for Space and Cosmic Evolution, Ehime University, 2-5 Bunkyo-cho, Matsuyama, Ehime 790-8577, Japan}

\author{Kodai Nobuhara}
\affil{Graduate School of Science and Engineering, Ehime University, Bunkyo-cho, Matsuyama, Ehime 790-8577, Japan}

\author{Chien-Hsiu Lee}
\affil{Subaru Telescope, National Astronomical Observatory of Japan, 650 North Aohoku Place, Hilo, HI 96720, USA}
\affil{National Optical Astronomy Observatory, 950 N. Cherry Ave., Tucson, AZ 85719, USA}

\author{Tomoki Morokuma}
\affil{Institute of Astronomy, Graduate School of Science, University of Tokyo, 2-21-1 Osawa, Mitaka, Tokyo 181-0015, Japan}

\author{Yoshiki Toba}
\affil{Department of Astronomy, Kyoto University, Kitashirakawa-Oiwake-cho, Sakyo-ku, Kyoto 606-8502, Japan}
\affiliation{Academia Sinica Institute of Astronomy and Astrophysics, P.O. Box 23-141, Taipei 10617, Taiwan}
\affil{Research Center for Space and Cosmic Evolution, Ehime University, 2-5 Bunkyo-cho, Matsuyama, Ehime 790-8577, Japan}

\author{Toshihiro Kawaguchi}
\affiliation{Department of Economics, Management and Information Science, Onomichi City University, Onomichi, Hiroshima 722-8506, Japan.}

\author{Akatoki Noboriguchi}
\affil{Graduate School of Science and Engineering, Ehime University, Bunkyo-cho, Matsuyama, Ehime 790-8577, Japan}

\begin{abstract}
We report the result of optical identifications of FIRST radio sources with
the Hyper Suprime-Cam Subaru Strategic Program survey (HSC-SSP). 
The positional cross-match within 1\arcsec\ between the FIRST and HSC-SSP catalogs ($i\lesssim 26$)
produced more than 3600 optical counterparts in the 156~deg$^2$ of the HSC-SSP field.
The matched counterparts account for more than 50\% of the FIRST sources in the search field, 
which substantially exceed previously reported fractions of SDSS counterparts 
($i\lesssim 22$) of $\sim 30$\%.
Among the matched sample, 9\% are optically unresolved sources such as radio-loud quasars.
The optically faint ($i>21$) radio galaxies (RGs) show that 
the fitting linear function of the 1.4~GHz source counts has a slope that 
is flatter than that of the bright RGs, 
while optically faint radio quasars show a slope steeper than 
that of bright radio quasars.
The optically faint RGs show a flat slope in the $i$-band number counts down to 24~mag, 
implying either less-massive or distant radio-active galactic nuclei (AGNs) beyond 24~mag.
The photometric redshift and the comparison of colors with the galaxy models show
that most of the matched RGs are distributed at redshifts from 0 to 1.5.
The optically faint sample includes the high radio-loudness sources 
that are not seen in the optically bright sample. 
Such sources are located at redshift $z>1$. 
This study gives $\sim 1500$ radio AGNs lying at the optically faint end
and high-redshift regime not probed by previous searches.
\end{abstract}


\section{Introduction} \label{sec:intro}

Radio galaxies (RGs), as well as radio-loud active galactic nuclei (AGNs) themselves,
will provide us with great opportunities for understanding 
galaxy formation and evolution. 
RGs prefer massive host galaxies (stellar mass of more than 10$^{11}\,M_{\odot}$; \citealt{Seymour2007}).
The powerful radio jets from RGs can drive AGN feedback to regulate 
star formation in massive host galaxies \citep{McNamara2005,Fabian2012,Morganti2013}.
Thus, RGs are thought to be a key population in the formation of massive galaxies. 
Moreover, high-$z$ RGs ($z>2$) could be a valuable beacon to find protoclusters,
because high-$z$ RGs tend to reside in overdensity regions \citep{McCarthy1993,Miley2008}.
In this paper, RGs are defined as radio-loud galaxies 
with rest-frame 1.4~GHz radio luminosities
of $>10^{24}$\,W\,Hz$^{-1}$ \citep{Tadhunter2016} and optically resolved morphologies, 
while radio-loud quasars are optically unresolved objects 
satisfying the same criterion as the radio luminosity.
We refer to both RGs and radio-loud quasars as radio AGNs.

The bright radio radiation of RGs is advantageous for probing high-$z$ AGNs.  
Such radio sources should be found by wide-area surveys in radio wavelengths
(NVSS, \citealt{Condon1998}; FIRST, \citealt{Becker1995,Helfand2015}; SUMSS, \citealt{Bock1999};
TGSS ADR, \citealt{Intema2017}), which follows the pioneering Cambridge Catalog series (e.g., \citealt{Edge1959}). 
It is greatly helpful to identify optical counterparts of the radio sources 
to investigate the properties of host galaxies and the optical emission of AGNs.

The Sloan Digital Sky Survey (SDSS; \citealt{York2000}) allowed identifications of 
the optical counterparts
of radio sources, taking advantage of its large sky coverage.
\citet{Ivezic2002} and subsequently \citet{Helfand2015} performed the positional cross-match 
between the SDSS and FIRST catalogs and
identified SDSS counterparts only for $\sim 30$\% of the FIRST radio core sources.  
Even though radio sources with an extended complex radio morphology,
which are a minority in all the FIRST sources ($\sim 10$\%; \citealt{Ivezic2002}),
were taken into consideration, 
the matching fraction of the SDSS objects with FIRST sources did not improve (e.g., \citealt{Ching2017}).
This means that more than half of all the FIRST sources do not have SDSS counterparts.
This is likely due to 
the shallow optical sensitivities
of SDSS imaging ($r<22.2$~mag).

In fact, a deep optical survey in the VVDS 1~deg$^2$ field showed that 
more than half of radio sources with a flux of $>1$~mJy in 1.4~GHz,
which is the same level as the FIRST sensitivity, match optically faint 
counterparts with $I > 22.9$~mag \citep{Ciliegi2005}.
In a 1.4~GHz VLA survey in the 1.3~deg$^2$ SXDF, 
\citet{Simpson2006, Simpson2012} found a few hundred optical counterparts 
of high-$z$ radio AGNs with $i^{\prime}<26$~mag up to $z\sim 4$.
Recently, \citet{Smolcic2017a,Smolcic2017b} performed a 3~GHz VLA deep imaging and
reported more than a thousand radio AGNs with $<26$~mag in the optical bands,
up to $z\sim 5$ in the 2.6~deg$^2$ COSMOS field.
\citet{Williams2018} performed an identification of radio sources in
the deep 150~MHz LOFAR observation with $I$-band-selected objects ($I < 24$),
and they reported $\sim 2000$ RGs up to $z\sim 2$ in 9.2~deg$^2$ of the Bo\"{o}tes field.
The previous deep optical surveys in the radio survey regions 
have successfully identified high-$z$ radio AGNs,
but the number of the radio AGNs, in particular for high radio luminosity sources, 
is limited to a few hundred due to their small survey areas of typically $< 10$~deg$^2$.
The Hyper Suprime-Cam (HSC) Subaru strategic program (SSP; \citealt{Aihara2017a}) 
reaches a limiting magnitude of 25.9 in the $i$-band in a 1400~deg$^2$ area,
thus providing a key combination for identifying rare and faint objects such as radio AGNs.

This paper presents a new exploration for radio AGNs using 
a combination of the wide and deep optical catalog of the Subaru HSC-SSP and
the wide-field radio catalog of FIRST.
This exploration is carried out by an ongoing project, 
WERGS (a Wide and Deep Exploration of RGs with Subaru HSC).
In this paper, we particularly focus on the optical and radio properties of RGs discovered in this project. 
We refer to radio loudness, $\mathcal{R}$, which is the measure of radio activity of AGNs.
We use radio loudness as an observable value, which is defined as the ratio of the 1.4~GHz flux to the HSC band flux.
Radio loudness is typically defined as a ratio of the 5~GHz flux to the $B$-band flux derived from AGN emission,
where AGNs and quasars with $\log\mathcal{R}>1$ are considered to be radio-loud \citep{Kellermann1989}.
Because the HSC band flux is assumed to be dominated by a host galaxy of an RG,
the radio loudness in this study could be underestimated. 
This paper is organized as follows.
We introduce the two catalog data sets in Section \ref{sec:data}. 
The method and statistical results of a cross-match between the catalogs 
are presented in Section \ref{sec:sample}.
The optical and radio properties of the entire HSC-FIRST radio AGN sample are shown in Section \ref{sec:result}.
The properties of optically faint RGs are discussed
in Section \ref{sec:discussion}. 
A summary is presented in Section \ref{sec:summary}.
Throughout this paper, we use the AB magnitude system
and a  $\Lambda$CDM cosmological model with 
$H_0 = 70~{\rm km~s^{-1}~Mpc^{-1}}$, $\Omega_{\rm M} = 0.27$, 
and $\Omega_{\rm \Lambda} = 0.73$.

\section{The Data}\label{sec:data}
This search for radio AGNs is based on the positional cross-match between 
the HSC-SSP optical source catalog and the FIRST radio source catalog.
In this section, the two source catalogs are described.

\subsection{Subaru HSC Survey}\label{sec:HSCdata}
The HSC-SSP is a multiband (five broadband filters, $g$, $r$, $i$, $z$, 
and $y$, and four narrowband filters) 
imaging survey with HSC  
(\citealt{Furusawa2018}; Kawanomoto et al.\ 2018, in preparation; \citealt{Komiyama2018, Miyazaki2018})
installed on the Subaru 8.2~m telescope.
Three complementary layers---the Wide, the Deep, and the UltraDeep---are included in the survey. 
These layers are different from each other in area and depth.

We utilized the broadband forced photometry catalogs of the early data products internally released as S16A \citep{Aihara2017b}, 
which were obtained by observations from 2014 March to 2016 January.
The typical seeing is $\sim 0\farcs6$ in the $i$-band.
The astrometric uncertainty is $\sim 40$~mas in RMS according to 
the comparison with the Pan-STARRS1 Surveys.
The forced photometry of each band image is performed using a position and shape parameters
determined in a reference band. 
The highest priority band of a reference band is the $i$-band, followed in order by $r$, $z$, $y$, and $g$
(see \citealt{Bosch2017} for more details).
We used the two extremes of the survey layers: 
the Wide layer originally covering 456~deg$^2$ with a $5\sigma$ depth for 
point sources of $i= 26.4$, and the UltraDeep layer of 4~deg$^2$ with 
a depth of $i= 27.0$ \citep{Aihara2017b}.
The saturation magnitudes for point sources are 18.4 and 18.7 in the $i$-band for the Wide layer and
the UltraDeep layer, respectively.
Using the two extreme layers, we investigate the dependence of the rate of optical identification
on the depth of optical imaging.
We do not use the Deep layer because the imaging depth of the S16A Deep layer data
is still as shallow as that of the S16A Wide layer due to an incomplete observation in the Deep layer.
The Wide layer is composed of six distinct areas, and the UltraDeep layer has two areas, 
the UD-COSMOS and the UD-SXDS.
Because the UD-SXDS area in the S16A data has 
a shallower depth (10--40\%)
than that of the UD-COSMOS, 
we use only the UD-COSMOS out of the UltraDeep layer.
The detailed properties and the data process of the early data products are described in \citet{Aihara2017b} and \citet{Bosch2017}.

We produced two clean HSC-SSP samples of the HSC-SSP Wide and UltraDeep layers,
from the original HSC-SSP forced catalogs 
in order to exclude detector edges with low sensitivities and junk sources from the original catalogs.
Here, we use cModel magnitudes from HSC-SSP photometry 
and corrected them for Galactic extinction \citep{Schlegel1998}.
First, we masked the edges of each area. 
Around the edges, the number of visits at an object position is much smaller than that in the central regions.
The coordinates of the consequent areas after masking the edges are shown in Table \ref{tab:HSCclean}.
Additionally, we removed coadded patches with poor modeling of the PSF due to extremely good seeings
(see section 5.8.4 in \citealt{Aihara2017b} for details).
A patch is the minimum unit of images for data processing. 
The removed patches have color offsets larger than 0.075~mag 
from the stellar sequence in the color-color planes of either 
$g-r$ and $r-i$, $r-i$ and $i-z$, or $i-z$ and $z-y$ \citep{Akiyama2017,He2017}.
We also removed subregions in two patches in the XMM area, where some cModel photometries are 
significantly overestimated compared with aperture photometries.

Then, we imposed the following criteria:
(i) the number of visits at an object position (\verb|countinputs|) 
is greater than or equal to 3 (5) for the $g$ and $r$ ($i$, $z$, and $y$) bands 
in the Wide layer to ensure an observing depth of more than a certain sensitivity,
and greater than or equal to 14, 26, and 42 for $g$ and $r$, $i$ and $z$, and $y$, respectively, in the UD-COSMOS;
(ii) the central $3\times 3$ pixels of an object are not affected by nearby bright objects 
(\verb|flags_pixel_bright_object_center=False|; see \citealt{Coupon2017} for details);
(iii) an object is not deblended further (\verb|deblend_nchild=0|);
(iv) an object is a unique source (\verb|detect_is_primary=True|);
(v) the central $3\times 3$ pixels of an object are not saturated 
(\verb|flags_pixel_saturated_center = False|)
and (vi) not affected by cosmic rays (\verb|flags_pixel_cr_center = False|);
and (vii) an object is not at the edge of a CCD or a coadded patch
(\verb|flags_pixel_edge = False|).
The following conditions were imposed only for the $r$, $i$, and $z$ bands:
(viii) there are no problems with measuring the cModel flux (\verb|cmodel_flux_flags = False|)
and (ix) centroid positions (\verb|centroid_sdss_flags = False|),
and (x) a signal-to-noise ratio (S/N) of an object is higher than 5 in all three bands.
As a result, 23,795,523 sources in a 154~deg$^2$ area and 
643,932 sources in a 1.78~deg$^2$ area are selected as the clean HSC-SSP samples 
of the Wide layer and the UD-COSMOS, respectively.

\begin{table}[t!]
\centering
\caption{Defined Coordinates (J2000.0) of Each HSC-SSP Wide Area in Our Survey} \label{tab:HSCclean}
\begin{tabular}{lll}
\tablewidth{0pt}
\hline
\hline
Field & \multicolumn{2}{c}{Coordinates\tablenotemark{\dag} (deg)} \\
        &  [$\alpha$(min), $\alpha$(max)] & [$\delta$(min), $\delta$(max)] \\
\hline
\decimals
GAMA15H  & [210.80, 224.90]& [-1.40,  1.08] \\
VVDS     & [331.20, 336.15]& [-0.45,  1.48] \\
GAMA09H  & [129.05, 135.45]& [-0.20,  2.26] \\
         & [135.45, 139.15]& [-1.15,  2.89] \\
         & [139.15, 140.40]& [-0.55,  2.89] \\
         & [138.64, 140.40]& [ 2.89,  4.45] \\
XMM      & [29.80, 38.53] & [-6.05, -2.55]   \\
HECTOMAP & [237.08, 247.64]& [42.40, 44.35] \\
WIDE12H  & [176.53, 182.51]& [-1.40,  1.07] \\
\hline
\end{tabular}

{\dag~ The coordinates of each area are represented by 
$\alpha{\rm (min)} \leq \alpha \leq \alpha{\rm (max)}$ 
and $\delta{\rm (min)} \leq \delta \leq \delta{\rm (max)}$.}
\end{table}

\subsection{FIRST Radio Survey}
The FIRST survey (the Faint Images of the Radio Sky at Twenty-cm; \citealt{Becker1995}) is 
a wide-field survey with the Very Large Array (VLA) at 1.4~GHz.
The spatial resolution is 5\farcs4, and 
the statistical astrometry error is 0\farcs2 for bright and compact sources \citep{White1997}.
The sky coverage is 10,575~deg$^2$, which nearly covers the full SDSS sky area.
The wide coverage and the high spatial resolution of FIRST are appropriate
for finding a large number of optical counterparts of radio sources.
The FIRST detection limit is 1~mJy in 5$\sigma$ sensitivity,
although only in a region at the equatorial strip
the limit is only 0.75~mJy.
Radio sources detected with a radio flux of more than the FIRST detection limit at a redshift $z\gtrsim0.5$
meet the definition of radio AGNs, 
where a radio AGN should have a 1.4~GHz luminosity
of $L_{\rm 1.4~GHz} > 10^{24}$~W~Hz$^{-1}$ \citep{Tadhunter2016}.

We use the final release catalog of FIRST \citep{Helfand2015}.
To produce a uniform sample, we use only sources with a radio flux of more than 1~mJy.
The FIRST catalog gives a probability $P(S)$ that a source is a spurious detection near a bright source.
We employed the probability and use only sources with $P(S) < 0.05$. 
Consequently, the FIRST clean samples in the unmasked regions of the HSC-SSP Wide layer 
and UD-COSMOS layer have 7072 and 106 sources, respectively.
See the Appendix for details about how to count FIRST sources in the unmasked HSC-SSP regions.

\section{HSC-FIRST Radio Sources}\label{sec:sample}
\subsection{Positional Matching}\label{sec:matching}
\begin{figure}[t]
\epsscale{0.9}
\plotone{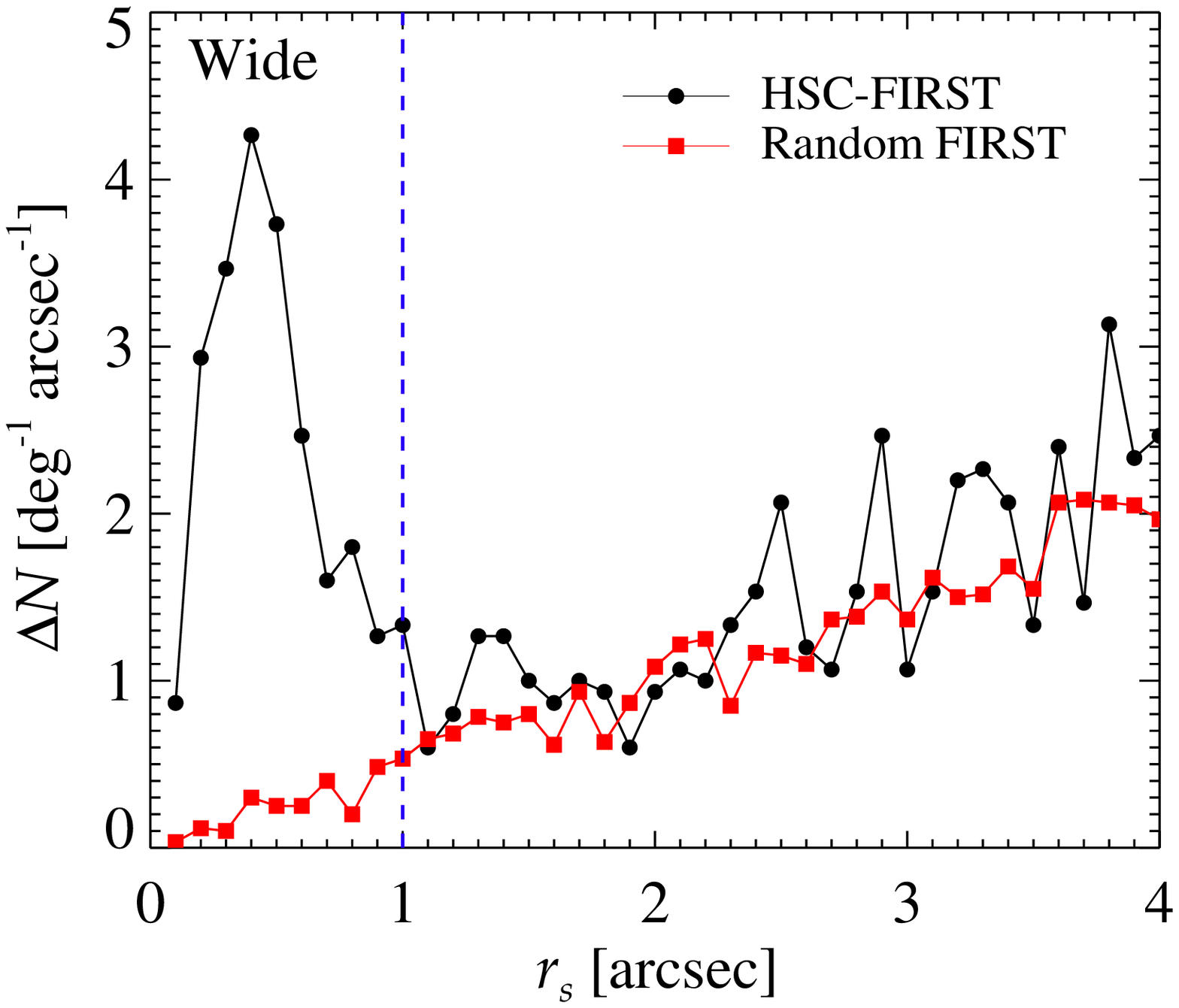}
\plotone{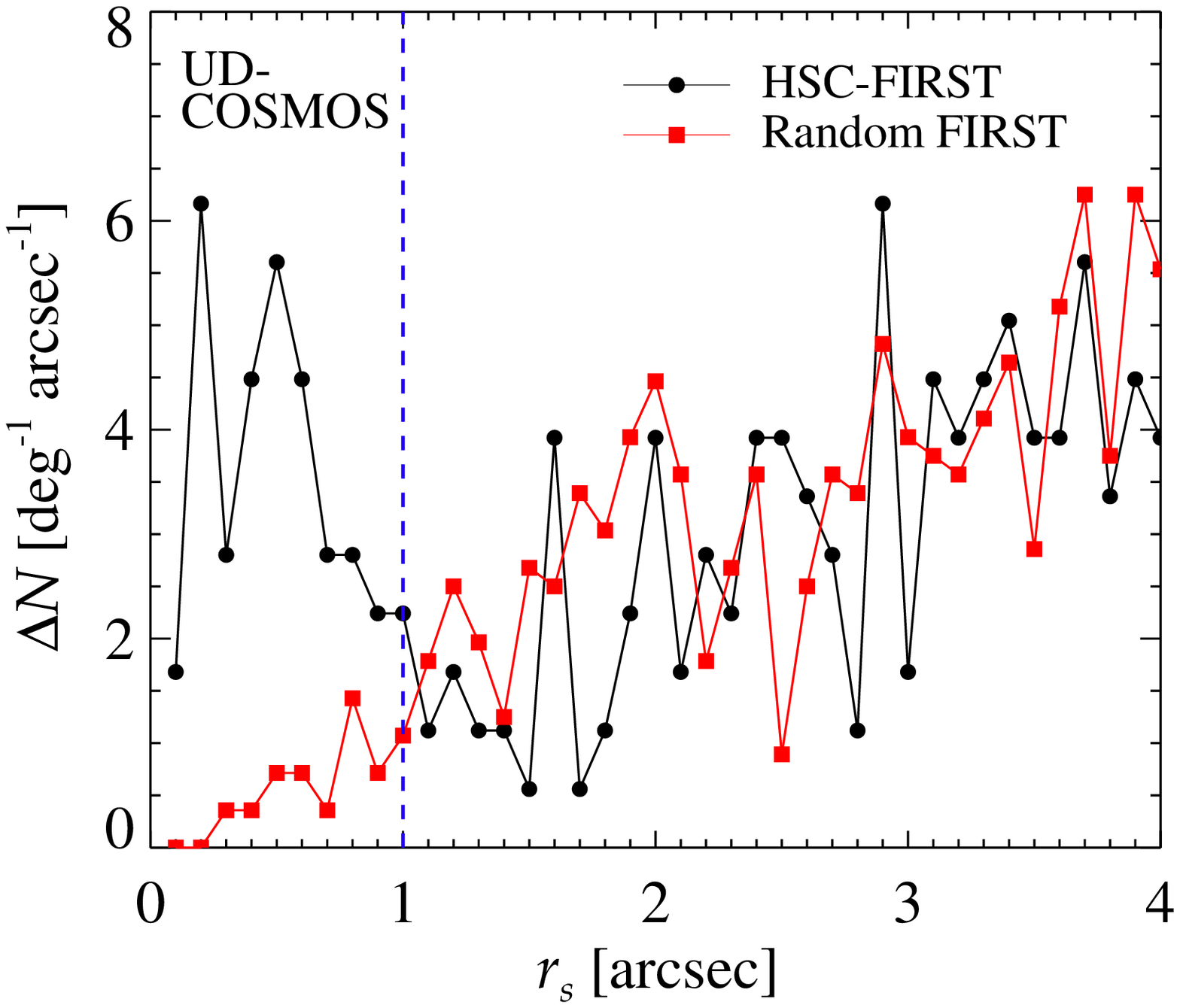}
    \caption{Differential distribution of the separation between the HSC-SSP and FIRST sources.
The red points represent the test results of the chance coincidence matches, 
which were performed using the mock FIRST catalogs with random positions.
The blue broken lines denote the adopted radius of 1\farcs0. 
({\it top}) the Wide layer, ({\it bottom}) UD-COSMOS.
\label{fig:search_r}}
\end{figure}

An HSC-SSP counterpart of a FIRST source is identified 
via nearest neighbor matching between
the HSC-SSP $i$-band and FIRST source positions.
The search radius $r_s$ is set to be 1\farcs0 for both cross-matching in the Wide and UD-COSMOS layers,
because $r_s=1\farcs0$ is an optimal condition according to an analysis of the completeness and
contamination of the cross-match as discussed below.
In Figure \ref{fig:search_r}, the differential distributions of the separation between 
the HSC-SSP and FIRST sources are compared with those of chance coincidence matches.
These examinations were performed in the central 15~deg$^2$ region of the GAMA09H Wide layer
and the central 0.7~deg$^2$ region of the UD-COSMOS.
The matches by chance coincidence are estimated 
from the average numbers of matches with the mock FIRST catalogs with random positions.
Each of the mock catalogs was generated by shifting the FIRST source positions from the original positions
to $+/-1\arcdeg$ or $+/-2\arcdeg$ along the R.A. direction for the Wide layer,
to $+/-0\fdg1$, $+/-0\fdg2$, $+/-0\fdg3$, or $+/-0\fdg4$ for the UD-COSMOS.
The numbers of matches have peaks at $0\farcs0 < r_s < 1\farcs0$ in both of the Wide and the UD-COSMOS layers.
Then, at $r_s > 1\farcs0$, 
the matches gradually increase with increasing radius, 
approaching the number of random matches.
Thus, the matched sources are dominated by chance coincidence at the radius of more than 1\farcs0.
It is worth mentioning that 31\% of the FIRST sources with HSC counterparts within $1\farcs0$
in the Wide layer
have multiple HSC sources (one to five sources) within the FIRST beam (2\farcs7 radius).
Among these multiple HSC sources, except for those HSC sources closed to a FIRST position,
97\% have separations larger than 1\farcs0, and thus can be attributed to the chance coincidence.

We did not apply any particular algorithm intended for complex and extended radio sources to our cross-match.
About only 10\% of the FIRST sources have such complex and extended morphology \citep{Ivezic2002,Ching2017}.
Therefore, our cross-matching method can miss radio AGNs without a radio core
and can misidentify radio lobes whose position matches an HSC-SSP source by chance.

\begin{deluxetable*}{crrrrrr}[t]
\tablecaption{Summary of the Cross-matching Results \label{tab:match_result}}
\tablecolumns{7}
\tablewidth{0pt}
\tablehead{
\colhead{HSC Layer} &
\colhead{HSC\tablenotemark{a}} &
\colhead{FIRST\tablenotemark{a}} & 
\colhead{Match\tablenotemark{a}} & 
\colhead{Matching Rate\tablenotemark{b}} & \colhead{Area (deg$^2$)}
}
\startdata
Wide & 23,795,523 & 7072 & 3579 & 0.506 & 154.09\\
UD-COSMOS & 643,932 & 106 & 63 & 0.594 & 1.78 \\
\enddata
\tablenotetext{a}{The number of sources within the same effective area 
specified in the last column.}
\tablenotetext{b}{The matching rate is defined by a ratio of 
the number of the matched sources to the number of the FIRST sources.}
\end{deluxetable*}

\subsection{Contamination and Completeness}
\begin{figure}[t]
\plotone{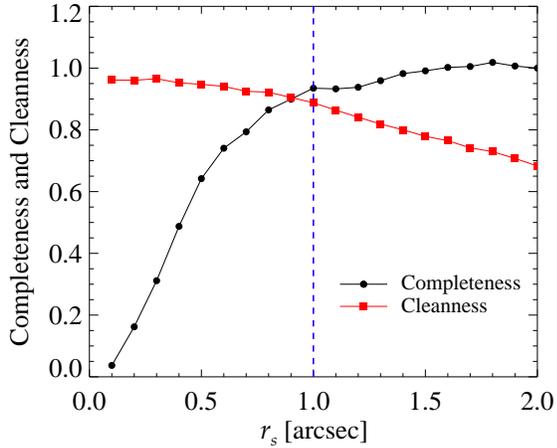}
\caption{Cumulative distribution of completeness (black) and cleanness (red) within each searching radius
in the Wide layer. 
The blue broken line indicates the adopted radius of 1\farcs0.
\label{fig:completeness}}
\end{figure}

Adopting a single search radius in the positional cross-match
allows us to estimate the
contamination and completeness of the source identifications
in the following way.
A total number of matches at $r_s<r_s^{\prime}$ is represented by $N_{\rm tot}$,
where $r_s^{\prime}$ is an adopted search radius.
The matches by chance coincidence at $r_s<r_s^{\prime}$, $N_{\rm cc}$, 
were estimated by the cross-match with the mock catalogs in the way mentioned in the previous section. 
The contamination is defined as the ratio of $N_{\rm cc}$ to $N_{\rm tot}$.
For convenience, we also defined a cleanness that is defined as ($1 -$ contamination).
A number obtained by subtracting $N_{\rm cc}$ from $N_{\rm tot}$ is 
represented by $N_{\rm A}$. 
We furthermore defined a number of matches, $N_{\rm B}$, 
which is a number obtained by subtracting matches by chance coincidence 
at $r_s^{\prime} < r_s < r_s^{\prime\prime}$ from the total number of matches 
in this radius range.
We set $r_s^{\prime\prime}$ of 2\farcs0, where the completeness is assumed to be 100\%.
Finally, the completeness is defined as a ratio of $N_{\rm A}$ to a sum of $N_{\rm A}$ and $N_{\rm B}$.
In Figure \ref{fig:completeness}, we show the cleanness and completeness 
as a function of $r_s$ for the Wide layer.
The choice of the searching radius proved to be a reasonable trade-off 
between completeness and cleanness of the sample. 
Indeed, at the adopted searching radius of $r_s \sim 1\farcs0$,
the contamination by chance
coincidence is estimated to be 14\%,
and the completeness is estimated to be 93\% in the Wide layer.
In the UD-COSMOS, while the completeness is not well determined 
due to the small number of matches, 
the contamination is 16\%.

The results of the cross-matches are summarized in Table \ref{tab:match_result}.
The cross-matches produced 3579 and 63 matched sources in the Wide and the UD-COMOS layers, respectively.
The matching rate, defined as the fraction of matched FIRST sources, is 51\% in the Wide layer.
The rate increases to 59\% in the UD-COSMOS.
Considering the above statistics, 
out of 3579 (63) matched HSC-FIRST sources in the Wide layer (UD-COSMOS), 
3078 (53) are true optical counterparts of the FIRST sources, 
and 501 (10) FIRST sources are likely to be random matches.
The estimated completeness implies that 
3310 FIRST sources have HSC-SSP counterparts in the Wide layer,
suggesting a completeness-corrected matching rate of 47\%.
These rates are higher than those of the previous studies using the SDSS
($\sim 30\%$; \citealt{Ivezic2002,Helfand2015}).
The high matching rates with the HSC-SSP data are further discussed 
and compared with other studies in Section \ref{sec:high_matching_rate}.

The matched HSC-FIRST samples of both the Wide and UC-COSMOS layers include no known pulsars
in the ATNF pulsar catalog compiling all of the published rotation-powered pulsars \citep{Manchester2005}.

\subsection{Extended Radio Sources}

\begin{figure}[t!]
\epsscale{1.15}
\plotone{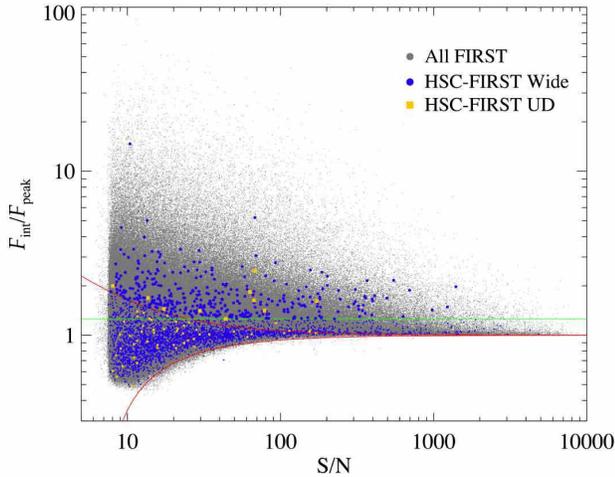}
\caption{Ratio of the total flux density to the peak flux density as a function of the radio S/N.
All of the FIRST sources are shown as gray dots.
The HSC-FIRST matched sources in the Wide layer are indicated by blue dots,
and those in the UD-COSMOS by yellow squares.
The sources above the upper red curve and the green line are considered to be extended radio sources (see main text).
 \label{fig:radio_extent}}
\end{figure}

In order to evaluate a number of sources with extended radio morphology in the HSC-FIRST matched samples,
we introduced a widely used size measure that is a ratio of the total integrated radio flux density
to the peak radio flux density, or $F_{\rm int}/F_{\rm peak}$ 
(e.g., \citealt{Ivezic2002,Schinnerer2007,Bondi2008,Smolcic2017a}).
The ratio $F_{\rm int}/F_{\rm peak}$ equals unity for a point source, 
while the ratio is larger than unity for an extended source or a resolved source.
In Figure \ref{fig:radio_extent}, we show the ratio as a function of the radio S/N.
All the FIRST sources (gray dots), the HSC-FIRST sources in the Wide (blue dots), 
and the UD-COSMOS (yellow squares) are shown. 
The $F_{\rm int}/F_{\rm peak}$ values of less than 1.0 are due to their low S/N.

We consider that sources simultaneously meeting the following two criteria should be extended radio sources.
The first one is a threshold line described by
\begin{equation}\label{eq:Ivezic_line}
\log{(F_{\rm int}/F_{\rm peak})} = 0.1,
\end{equation}
which was empirically determined for the FIRST-SDSS study by \citet{Ivezic2002},
and sources above the line were classified as radio extended sources.
This is, however, ineffective for the low-S/N sources.
For a second criterion, 
we determined the lower envelope (the lower red curve) of the data points in Figure \ref{fig:radio_extent} 
following \citet{Schinnerer2007}.
The envelope contains 99\% of all the FIRST sources below the $F_{\rm int}/F_{\rm peak}=1$ line
and was mirrored above $F_{\rm int}/F_{\rm peak}=1$.
The mirrored upper curve (the upper red curve) is represented as
\begin{equation}\label{eq:VLA_curve}
F_{\rm int}/F_{\rm peak} = 1+6.5\times(F_{\rm peak}/{\rm RMS})^{-1},
\end{equation}
where $F_{\rm peak}/{\rm RMS}$ is the S/N, and rms is a local rms noise in the FIRST catalog.
Finally, we considered that the sources located above 
both the line of Equation \ref{eq:Ivezic_line} and the curve of Equation \ref{eq:VLA_curve}
should be radio extended sources in our samples.

Consequently, 10\% (374 sources) of the HSC-FIRST Wide sample 
and 16\% (10 sources) of the HSC-FIRST UD-COSMOS sample are selected to be the radio extended sources.
These extended sources are likely radio lobes matched by chance, 
FR-I RGs \citep{Fanaroff1974}, RGs with core--jet/core--halo structures,
or nearby star-forming galaxies.
These fractions are thus upper limits of the numbers of chance coincidence of radio lobes,
and the upper limits are compatible with the contaminations estimated in Section \ref{sec:matching}.

\subsection{Optically Unresolved Sources}\label{sec:optical_morphology}

The optical morphologies on HSC images were examined to separate 
optically extended (resolved) sources and stellar-like unresolved objects
using the procedure described in \citet{Akiyama2017} 
for distinguishing RGs and radio-loud quasars.
We used a ratio of a second-order adaptive moment to that of a point spread function at a source position. 
Because $i$-band images were taken under better conditions than the other bands,
we use $i$-band images in this examination.
The criteria for selecting stellar objects are
\begin{eqnarray}
\verb|ishape_hsm_moments_11| / \verb|ishape_hsm_psfmoments_11|&& \nonumber \\
 < 1.1,\qquad&&\\
\verb|ishape_hsm_moments_22| / \verb|ishape_hsm_psfmoments_22|&& \nonumber \\
 < 1.1.\qquad&&
\end{eqnarray}
The suffixes of \verb|11| and \verb|22| represent each perpendicular axis in an image.
We adopted the adaptive moments of the \verb|HsmMoments| algorithm \citep{Hirata2003}.
For 14 sources with failed \verb|HsmMoments| measurements, 
the alternative adaptive moments of the \verb|SdssShape| algorithm \citep{Lupton2001} were adopted.
According to \citet{Akiyama2017},
the completeness of classified stellar objects in this classification decreases 
and the contamination from extended objects increases with increasing magnitude
due to the misclassifications.
At $i=23$, the completeness is 80--90\% and the contamination is less than 5\%, 
while at $i>24$, the completeness decreases down to $<70$\% 
and the contamination is more than 40\%.
Therefore, we consider that this classification of stellar objects 
is reliable for sources brighter than $i=24$~mag.

According to the classification, 
we found 326 and 2 stellar objects from the HSC-FIRST Wide sample and the UD-COSMOS sample, respectively.
These correspond to 9.1\% and 3.2\% of each sample, respectively.
For brevity, we refer to such stellar objects as quasars and to optically resolved sources as RGs or galaxies,
despite the recognition of exceptions to and uncertainties of the classification.

\subsection{Astrometry}

\begin{figure}[t!]
\plotone{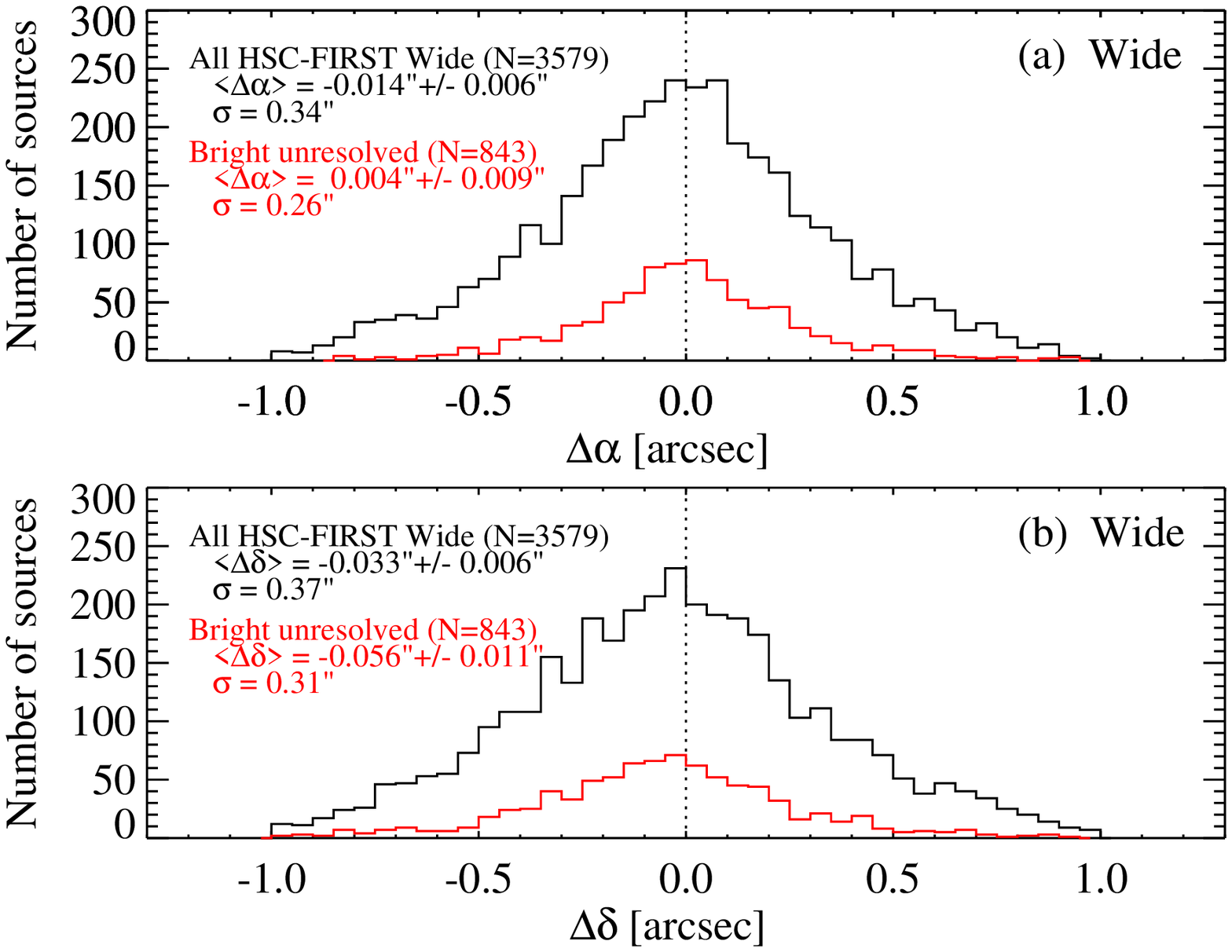}
\plotone{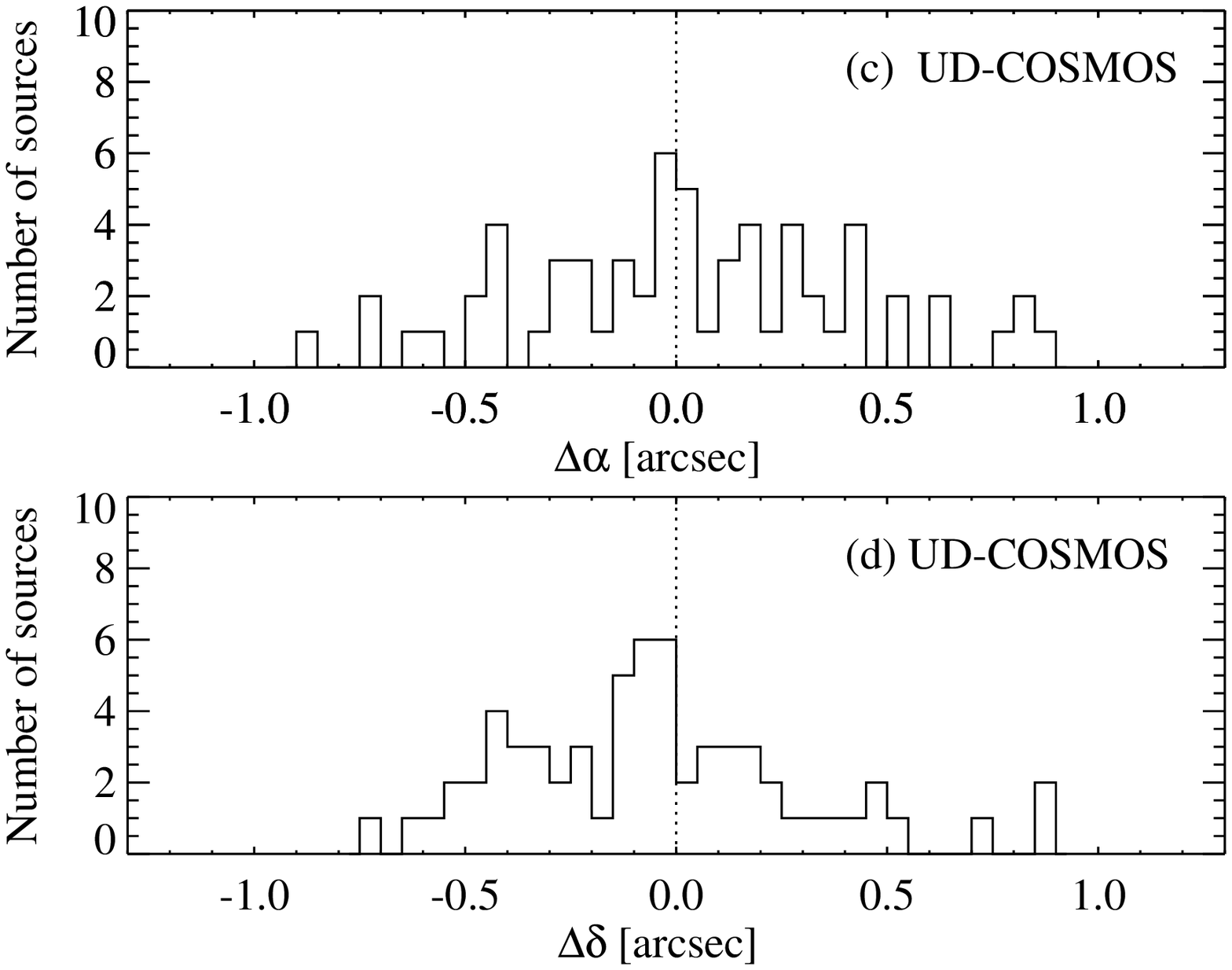}
\caption{Separations ($\Delta \alpha$ and $\Delta \delta$) between an HSC position and a FIRST position
in the Wide HSC-FIRST sample (the upper panels)
and the UD-COSMOS sample (the lower). 
In the upper panels, the separations for all sources in the HSC-FIRST sample 
and the subsample of optically unresolved sources with $F_{\rm peak}>3$~mJy and $i<24$
are shown as black and red lines, respectively.
\label{fig:separation}}
\end{figure}

The large size of our sample provides us with an opportunity 
to inspect the astrometry between the HSC-SSP and FIRST sources.
In Figure \ref{fig:separation}, we show the distributions of 
the residuals between the source positions of HSC-SSP and FIRST.
The residuals along R.A. and decl. are
a FIRST position minus an HSC-SSP position.
We mention only results of the Wide layer because the UD-COSMOS sample 
is statistically too small to discuss the astrometry,
but we just show the results of the UD-COSMOS sample.

The standard deviations of all the 3579 HSC-FIRST Wide sample 
are 0\farcs34 in R.A. and 0\farcs37 in decl.
When we use only a bright and unresolved subsample whose 
sources are unresolved in radio morphology 
and have a radio peak flux of $>3$~mJy and $i$ magnitude of $<24$, 
the standard deviations slightly moderate down to 0\farcs26 in R.A. 
and 0\farcs31 in decl.
These uncertainties are still large compared with
the positional uncertainty of the HSC-SSP catalog
($0\farcs04$ against Pan-STARRS1; \citealt{Aihara2017b}), 
and slightly larger than the uncertainty of the FIRST catalog 
($0\farcs2$ for bright and compact sources; \citealt{White1997}).
This is likely due to the galaxies composing almost the entire HSC-FIRST sample; 
we do not limit the sample for this inspection only to optically unresolved objects 
because of its quite small sample size. 

We found a mean offset in R.A. of $-0\farcs014\pm 0\farcs006$ 
in all of the HSC-FIRST Wide sources.
However, when we used only the the bright and unresolved subsample, 
this systematic offset becomes $0\farcs004\pm 0\farcs009$ 
and statistically not significant.
On the other hand, 
the mean offset in decl. is $-0\farcs033\pm 0\farcs006$ 
in the entire HSC-FIRST Wide sample.
This offset is still significant and is $-0\farcs056\pm 0\farcs011$, 
even if we use the bright and unresolved subsample.
Therefore, there is a systematic offset in decl. of the HSC-SSP 
sources with respect to the FIRST position.
Although this offset is significant,
it is much smaller than the positional uncertainty of the FIRST catalog ($0\farcs2$),
thus this offset has a negligible impact on the cross-match done in this work.

It may be valuable to compare the previous results of the FIRST sources with the SDSS counterparts
studied by \citet{Helfand2015}.
They reported that there are systematic offsets of $\sim 0\farcs02$ in R.A. and $\sim 0\farcs01$ in decl. 
of the FIRST positions with respect to both the SDSS sources 
and the FIRST sources that have been 
taken with the Karl G. Jansky Very Large Array (JVLA), which is a major upgrade of VLA.
Hence, taking this previous result into consideration, there might also be offsets 
between HSC-SSP and SDSS.
From this simple comparison, we inferred systematic offsets of $\sim 0\farcs02$ in R.A. 
and $\sim -0\farcs07$ in decl. from SDSS to HSC-SSP.

\section{Radio and Optical Properties of HSC RGs}\label{sec:result}
In this section, radio and optical properties of the HSC-FIRST matched samples are described. 
In particular, we investigate differences between RGs and radio-loud quasars, 
which were classified in Section \ref{sec:optical_morphology}.
Moreover, we focus on differences between two sub-samples:
the optically-faint HSC-FIRST sources (HSC-level), 
for which an optical counterpart is detected with HSC-SSP but is not with SDSS, 
and the optically-bright HSC-FIRST sources (SDSS-level),
for which an optical counterpart is detected both with HSC-SSP and SDSS.

\begin{figure}[t!]
\epsscale{1.15}
\plotone{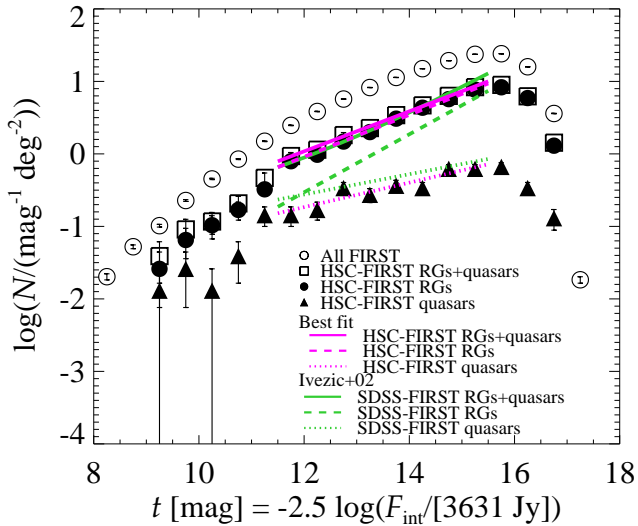}
\caption{
1.4~GHz radio source counts of the HSC-FIRST Wide sample (square).
The sample is separated into the HSC-FIRST RGs (filled circle) and
the HSC-FIRST quasars (triangle).
For comparison, all of the FIRST sources (open circle) are also shown.
The 1.4~GHz flux density is converted to the equivalent AB magnitude, $t$.
The Poisson errors in the data are indicated by vertical bars.
The best fits for three HSC-FIRST samples in $11.5 < t < 15.5$ are overplotted (magenta lines).
For comparisons, the distributions of the SDSS-FIRST samples \citep{Ivezic2002} 
are shown as their best-fit lines (green lines).
The SDSS-FIRST RG and quasar samples are morphologically classified subsets of optically resolved
and unresolved sources out of the SDSS-FIRST sample, respectively.
Both samples are limited to be brighter than 21 in $r^*$ and have robust classifications.
The SDSS-FIRST all sample is a combination of both the SDSS-FIRST RG and quasar sources, 
multiplied by 1.56 to retrieve the extracted sources by the $r^*<21$ condition
in the analysis by \citet{Ivezic2002}.
64\% sources meet their criteria.
\label{fig:radio_sourcecounts}}
\end{figure}

\begin{figure*}[th!]
\gridline{\fig{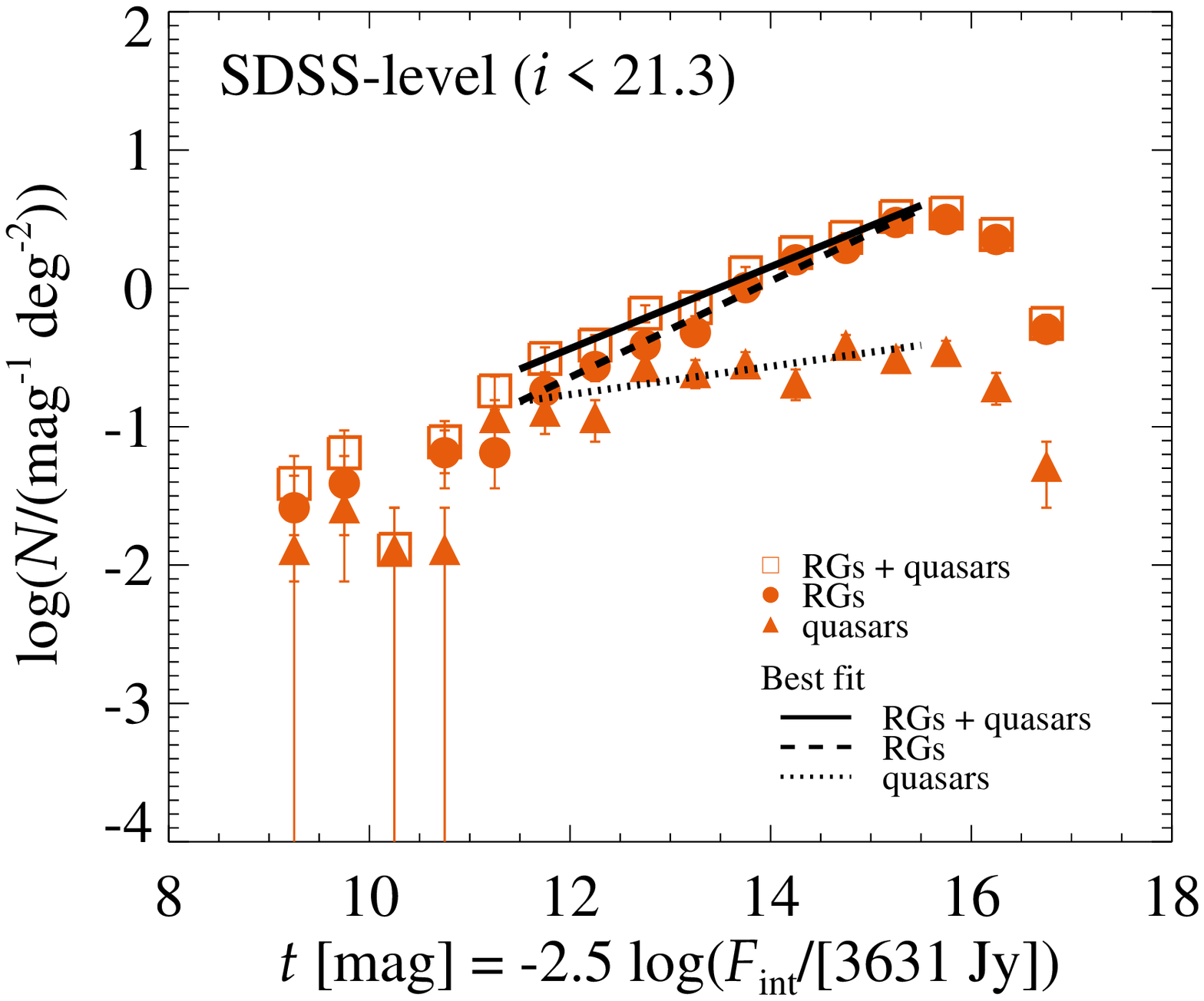}{0.32\textwidth}{}
          \fig{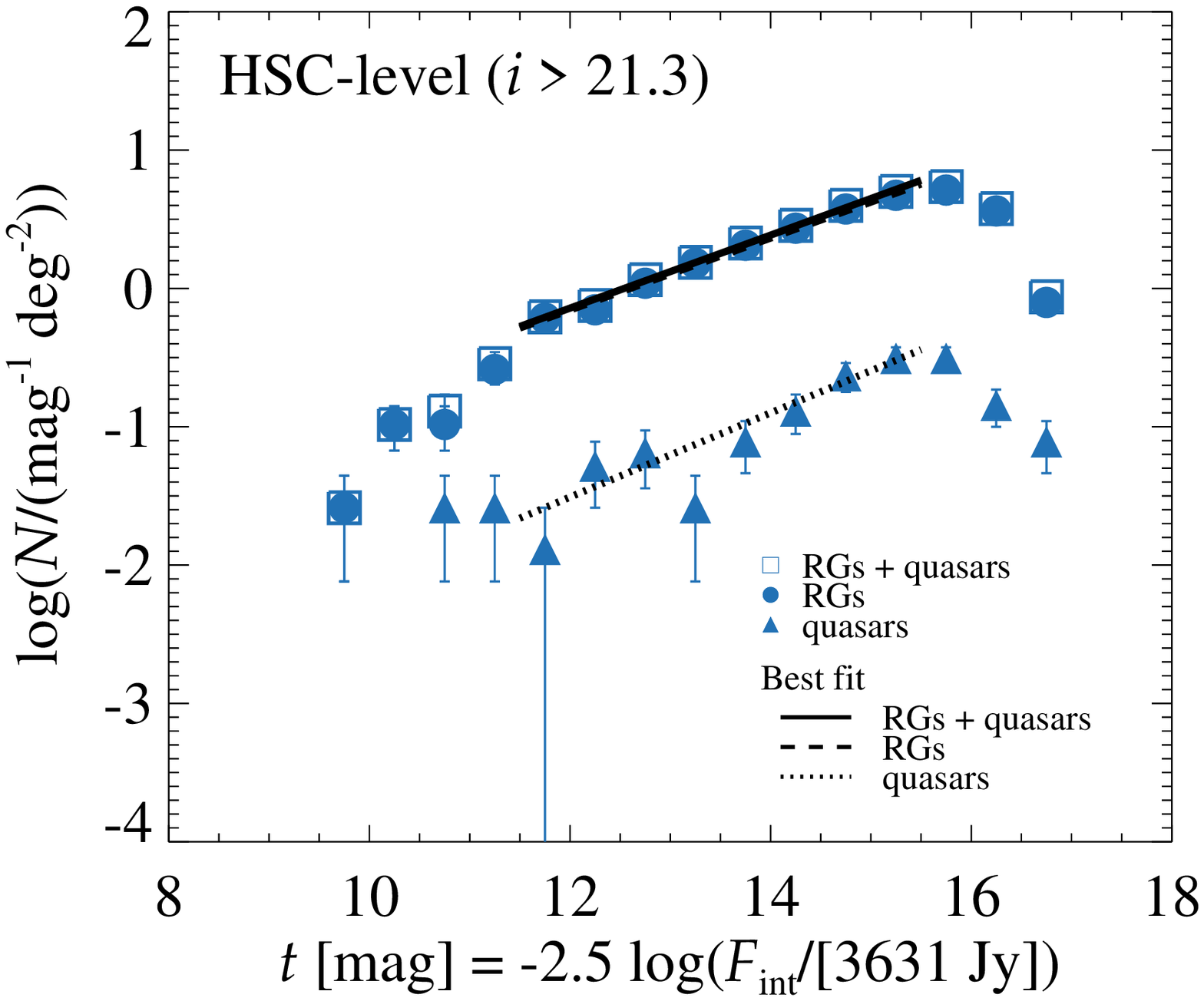}{0.32\textwidth}{}
          \fig{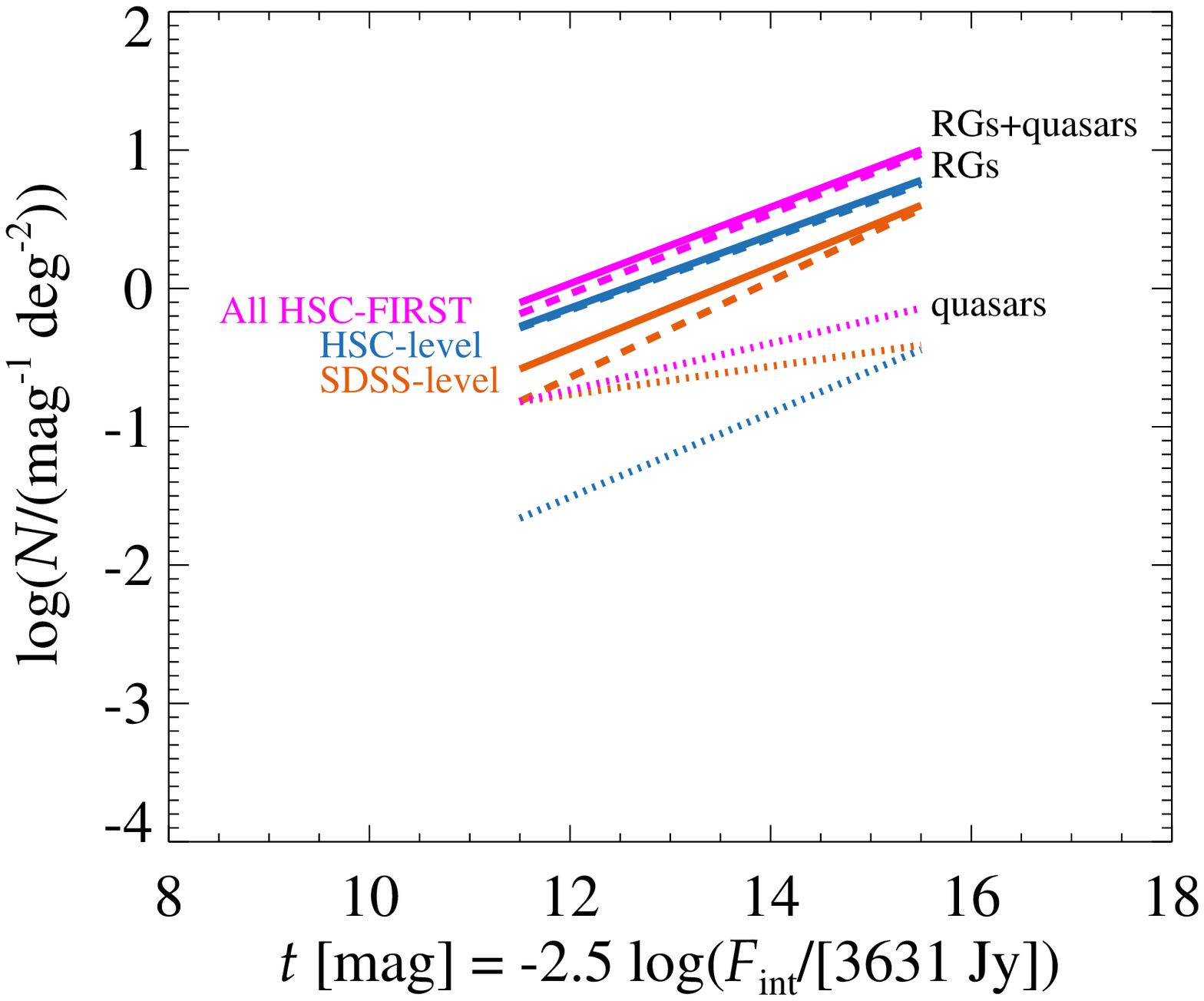}{0.32\textwidth}{}
          }
\vspace{-5ex}
\caption{
1.4~GHz radio source counts of the optically bright subsample, SDSS-level (left panel), and 
the optically faint subsample, HSC-level (middle panel). See the text for details of the subsamples.
Each subsample is further separated into RGs (circle) and quasars (triangle), which are
classified by optical morphologies.
The best-fit lines of each group are shown.
The fit lines of the subsamples (orange lines for the SDSS level, blue lines for the HSC level)
and those for all of the HSC-FIRST sources (magenta lines) are compared to each other (right panel).
The fit lines for the RGs (broken lines) and quasars (dotted lines) of each sample are also indicated.
\label{fig:radio_sourcecounts_SDSS_HSC}}
\vspace{5ex}
\end{figure*}

\subsection{Radio Source Counts}
Radio properties of the Wide sample are investigated using 1.4~GHz radio source counts in Figure \ref{fig:radio_sourcecounts}.
Although radio source counts are often examined on a plane of radio flux and $(dN/dS)/S^{-2.5}$,
we use another plane of radio AB magnitude and $N$ to directly compare with the results of \citet{Ivezic2002}.
Here, the radio AB magnitude is expressed as
\begin{equation}
t~{\rm [AB~mag]} = -2.5 \log{(F_{\rm int}/3631~{\rm Jy})}.
\end{equation}
A flux density of 1~mJy is equivalent to $t=16.4$.
In Figure \ref{fig:radio_sourcecounts}, 
the source counts of the HSC-FIRST Wide sample, the Wide RG sample, and the Wide quasar sample are shown.
Their distributions at a range of $11.5<t<15.5$ are fitted with a linear function by minimizing the chi-square statistic.
The magnitude range for fitting is the same one as the analysis by \citet{Ivezic2002}, whose
results are also shown in Figure \ref{fig:radio_sourcecounts}.
The parameters of the best fits are summarized in Table \ref{tab:fit_radiosourcecount}.

The HSC-FIRST Wide sample (square) is roughly consistent with the FIRST sources 
whose SDSS counterparts are identified by \citeauthor{Ivezic2002} (2002; the SDSS-FIRST sample, green solid line).
This is trivial because the $\sim 20$\% improvement of the matching rate in this study corresponds to
an increase of only $\sim 0.2$ dex in $\log{N/\rm (mag^{-1}\,deg^{-2})}$ and 
a fact that FIRST sources without SDSS counterparts draw a similar slope 
to those with SDSS counterparts \citep{Ivezic2002}.

The HSC-FIRST quasars have a somewhat lower number than the SDSS-FIRST quasars, 
while the HSC-FIRST RGs increase in number, in particular, at the bright end.
To investigate where this discrepancy stems from,  
we separate the HSC-FIRST Wide sample into an SDSS-level subsample with $i<21.3$
and an HSC-level subsample with $i>21.3$.
The $i$ magnitude threshold corresponds to the limiting magnitude of the SDSS
survey, where the catalog is 95\% complete for point sources \citep{Stoughton2002}.
The 1.4~GHz radio source counts of the two subsamples and the fitting results are shown 
in Figure \ref{fig:radio_sourcecounts_SDSS_HSC} and Table \ref{tab:fit_radiosourcecount}.
The slope of the HSC-level RGs is flatter than that of the SDSS-level RGs, 
while the slope of the HSC-level quasars is steeper than that of the SDSS-level quasars.
Therefore, the increased number of the HSC-FIRST RGs is likely due to 
the different dependences of the optically faint RGs and quasars
from the optically bright sources on radio flux.

A fraction of quasars out of the SDSS-level subsample increases with increasing radio flux,
from $\sim 20$\% at the faint end to $\sim 60$\% at the bright end.
On the other hand, in the HSC-level subsample, the fraction does not strongly depend on radio flux.
The fraction is approximately 5\% over the radio flux range for fitting.
The results in this subsection hold even if the HSC-level subsample 
is limited to sources with $i<24$.

\begin{table}[b!]
\centering
\caption{Fitting Results of 1.4~GHz Radio Source Counts} \label{tab:fit_radiosourcecount}
\begin{tabular}{ccrr}
\tablewidth{0pt}
\hline\hline
     &             &  \multicolumn{1}{c}{$a$}    & \multicolumn{1}{c}{$b$} \\
\hline
\decimals
All  & RGs         & $-3.51$ (0.29) & 0.289 (0.020) \\
     & quasars     & $-2.76$ (0.78) & 0.169 (0.056) \\
     & RGs+quasars & $-3.28$ (0.27) & 0.276 (0.019) \\
\hline   
SDSS-level  & RGs         & $-4.82$ (0.54) & 0.348 (0.038)\\
($i<21.3$)  & quasars     & $-1.99$ (0.94) & 0.102 (0.067)\\
            & RGs+quasars & $-3.99$ (0.45) & 0.296 (0.032)\\
\hline
HSC-level  & RGs         & $-3.28$ (0.35) & 0.260 (0.024)\\
($i>21.3$) & quasars     & $-5.16$ (1.79) & 0.305 (0.123)\\
           & RGs+quasars & $-3.31$ (0.34) & 0.264 (0.024)\\
\hline
\multicolumn{4}{p{0.45\textwidth}}{\textsc{Note}--- The best-fit parameters of the radio source counts 
in Figure \ref{fig:radio_sourcecounts} and \ref{fig:radio_sourcecounts_SDSS_HSC}.
The fitting function is $\log{(N/{\rm [mag^{-1}\,deg^{-2}]})} = a + b~t\,\rm [AB~mag]$.
The numbers in the parentheses represent $1\sigma$ uncertainties in the fitting results.
}
\end{tabular}
\end{table}

\subsection{Optical Number Counts}

\begin{figure}[t!]
\epsscale{1.16}
\plotone{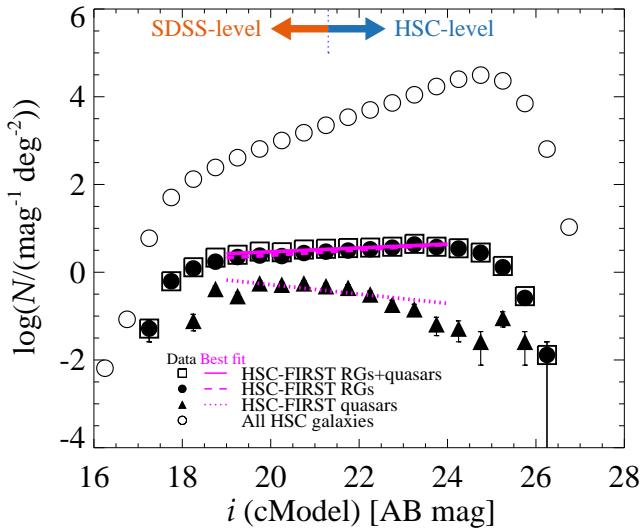}
\caption{
Optical HSC $i$-band number counts of the HSC-FIRST samples.
The symbols are the same as Figure \ref{fig:radio_sourcecounts}
but the circle indicates all of the HSC galaxies. 
The best fits are also shown by magenta lines for the HSC-FIRST Wide,
Wide RGs, and Wide quasar samples.
\label{fig:i_numbercounts}}
\end{figure}

Thanks to the depth of the HSC-SSP data, 
this study enables us to investigate optical properties of radio AGNs
down to more than 3~mag fainter in brightness than the previous SDSS studies.
The $i$-band number counts of the HSC-FIRST Wide sample are shown in Figure \ref{fig:i_numbercounts}.
A fall-off of the HSC-FIRST Wide sample at the bright end ($i\lesssim 18$)
is likely due to the saturation effect of the HSC-SSP data. 
A dearth of the HSC-FIRST Wide quasars at the bright end is likely due to 
not only the saturation but also the rareness of radio-loud quasars.
The SDSS radio-loud quasars are actually identified even at $i\lesssim 18$ \citep{White2000,Ivezic2002}.
The $i$-band number counts of the SDSS radio-loud quasars 
overlap with those of the HSC-FIRST quasars around 18~mag (see Figure~7 in \citealt{Ivezic2002}).
On the other hand, at the faint end at $i\gtrsim 25$, 
the number counts of the HSC-FIRST sources fall. 
The magnitude of this fall-off coincide with that of a fall-off
for the HSC galaxy sample. 
Both fall-offs are considered to be due to the effect 
of the detection incompleteness of the HSC-SSP Wide data. 
It is presumed that the detection completeness is close to 100\% down to brighter than $\sim 24$~mag
because the $i$-band sources were detected with an SN ratio of more than $\sim 50$ at $\lesssim 24$~mag.
The number counts of the HSC-FIRST sources, which are not damped down to the detection completeness magnitude,
implies a possibility that there are optically much fainter radio AGNs.

The number counts were fitted by a linear function within the $19<i<24$ range 
(magenta lines in Figure \ref{fig:i_numbercounts}).
The flat slope of the best fit for the HSC-FIRST RGs means that their number counts
do not strongly depend on the apparent optical magnitude.
The flat slope also indicates a decreasing fraction of the optically faint RGs 
relative to the HSC galaxies with increasing apparent magnitude: from 0.3\% at $i=19$ 
to 0.01\% at $i=24$.
The small fraction of the faint RGs might imply a small number of distant RGs 
and/or RGs with a less massive host galaxy.
For the HSC-FIRST quasars, there is a turnover at $\sim 21$~mag. 
Thus, the linear function is not well fitted to the data.
The fraction to the HSC-FIRST sources drops off from 11\% to $< 2\%$ over the turnover magnitude.
This turnover appears at $\sim 21$~mag, and therefore,
the turnover is not due to the detection completeness of HSC-SSP.
The FIRST detection limit, in part, may cause the turnover 
because the quasar sample misses optically faint radio-loud quasars
with a lower radio flux than the FIRST detection limit. 
However, a similar turnover is not seen in the RG sample, 
although the effect of the FIRST detection limit should also impinge on the RG sample.
This contrast means that the dependence of the number of quasars on optical magnitude
is different from RGs in the optically faint regime. 
\citet{Jiang2007} found that the radio-loud fraction of quasars decreases 
with decreasing luminosity and increasing redshift
although their quasar sample was basically brighter than $i=18.9$.
The turnover in our radio-loud quasar sample, therefore, may be partially 
attributed to a small number of low-luminosity radio-loud quasars and/or 
distant radio-loud quasars.

\begin{table}[b!]
\centering
\caption{Fitting Results of the $i$-band Number Counts} \label{tab:fit_iND}
\begin{tabular}{p{10em}rr}
\tablewidth{0pt}
\hline\hline
          &  \multicolumn{1}{c}{$a$}    & \multicolumn{1}{c}{$b$} \\
\hline
\decimals     
RGs         & $-0.846$ (0.276) & 0.0617 (0.0127)\\
quasars     & $1.85^{a}$ (0.99) & $-0.106^{a}$ (0.047)\\
RGs+quasars & $-0.406$ (0.261) & 0.0436 (0.0120)\\
\hline
\multicolumn{3}{p{0.45\textwidth}}{\textsc{Note}--- The best-fit parameters of 
the $i$-band number counts in Figure \ref{fig:i_numbercounts}.
The fitting function is $\log{(N/{\rm [mag^{-1}\,deg^{-2}]})} = a + b~i\,\rm [AB~mag]$.
The numbers in the parentheses represent $1\sigma$ uncertainties in the fitting results.

$^a$~The statistical significance of the fitting result for the quasars is apparently low.
}
\end{tabular}
\end{table}

\subsection{Radio Loudness}

\begin{figure*}[t!]
\epsscale{0.8}
\plotone{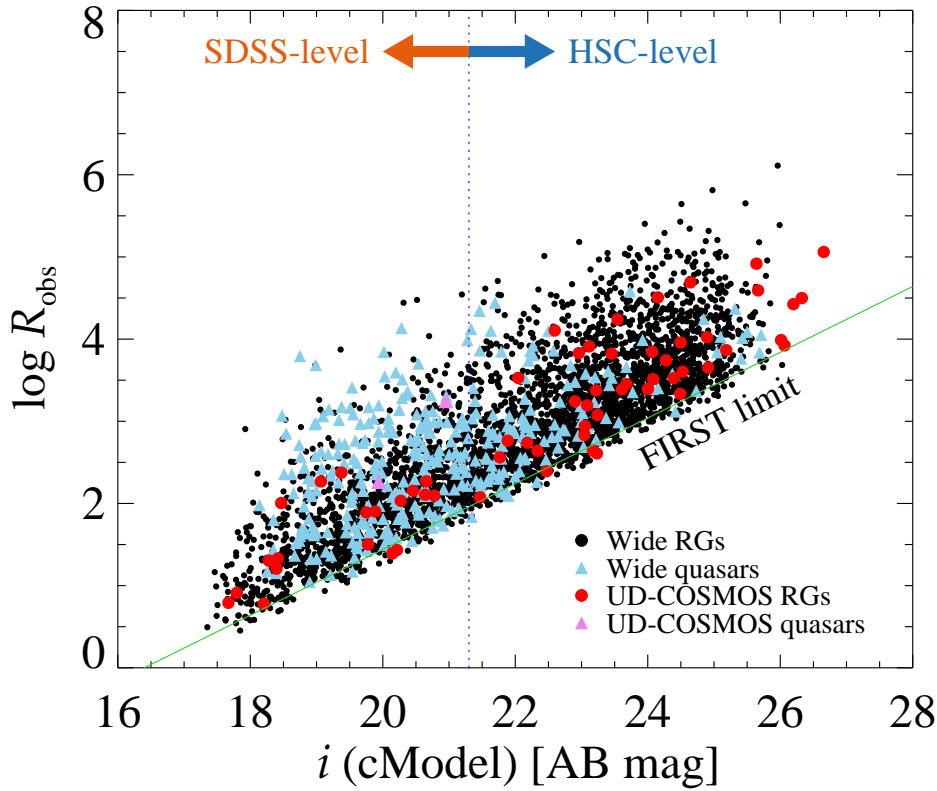}
\caption{
Logarithmic radio loudness $\log \mathcal{R}_{\rm obs}$ at the observed frame 
as a function of HSC $i$-band magnitude.
The RGs and quasars in the Wide sample are indicated by black circles and light blue triangles, respectively.
Of the UD-COSMOS sample, the RGs and quasars are shown as red circles and violet triangles, respectively.
The green line indicates the detection limit of FIRST point sources.
The blue dotted line is the border separating the SDSS-level and HSC-level samples by their $i$-band magnitude.
\label{fig:radioloudness}}
\vspace{4mm}
\end{figure*}

Radio loudness at the observed frame, $\mathcal{R}_{\rm obs}$, 
is calculated to examine the radio activities 
of the HSC-FIRST RGs and quasars.
According to \citet{Ivezic2002},
$\mathcal{R}_{\rm obs}$ is defined as a ratio of the 1.4~GHz flux and $i$-band flux 
at the observed frames without $k$-corrections.
The $k$-corrected radio loudness is discussed later (Section \ref{sec:discussion_properties}).
The logarithmic $\mathcal{R}_{\rm obs}$ is described as
\begin{equation}
\log \mathcal{R}_{\rm obs} = \log(F_{\rm int}/F_i) = 0.4\,(i - t).
\end{equation}
As mentioned in the introduction, 
the radio loudness could be underestimated for RGs,
because the stellar light of a host galaxy is assumed to be dominant in the $i$-band flux.

The $\mathcal{R}_{\rm obs}$ of the HSC-FIRST Wide and UD-COSMOS samples are shown in Figure \ref{fig:radioloudness}.
The HSC-FIRST samples obviously include higher $\mathcal{R}_{\rm obs}$ sources 
than the SDSS-level ones owing to the deep optical observations of HSC-SSP as we have expected. 
In particular, radio AGNs with more than $\log \mathcal{R}_{\rm obs} = 3$ 
substantially increase in number.
We note the $\mathcal{R}_{\rm obs}$ could be actually higher 
due to the underestimate by the contamination from host galaxies.
Moreover, the $\mathcal{R}_{\rm obs}$ of the HSC-level sample is biased toward 
high values due to FIRST's detection limit as illustrated by the green line in Figure \ref{fig:radioloudness}.
At the $i$-band magnitude range of the SDSS level, the quasars abound in the high $\mathcal{R}_{\rm obs}$ regime,
whereas the RGs with high $\mathcal{R}_{\rm obs}$ are at the HSC-level range.
The UD-COSMOS sample, as expected, includes optically fainter radio AGNs ($i\gtrsim 26$) than the Wide sample.
However, despite the deeper depth of UD-COSMOS, the highest $\mathcal{R}_{\rm obs}$ radio AGNs
are identified in the Wide layer. 
This means that wide optical surveys have an advantage over deeper but narrower surveys
in finding rare AGNs such as the radio-loudest AGNs.

\section{Discussion}\label{sec:discussion}
\subsection{Properties of HSC RGs as a Function of Photometric Redshift}
We have successfully identified a large number of 
the optically faint optical counterparts of FIRST radio sources
thanks to the wide and deep HSC-SSP survey.
We reported in the previous section that these optically faint FIRST sources show apparently high radio loudness.
In this section, we further investigate the optical and radio properties of the optically faint FIRST sources
in order to understand their nature.
We use the photometric redshift (photo-$z$ or $z_{\rm photo}$) of the sample
but quasars have large uncertainties in photo-$z$;
therefore we address only RGs other than quasars.

\subsubsection{Color--Color}\label{sec:colorcolor}
\begin{figure}[t!]
\epsscale{1.15}
\plotone{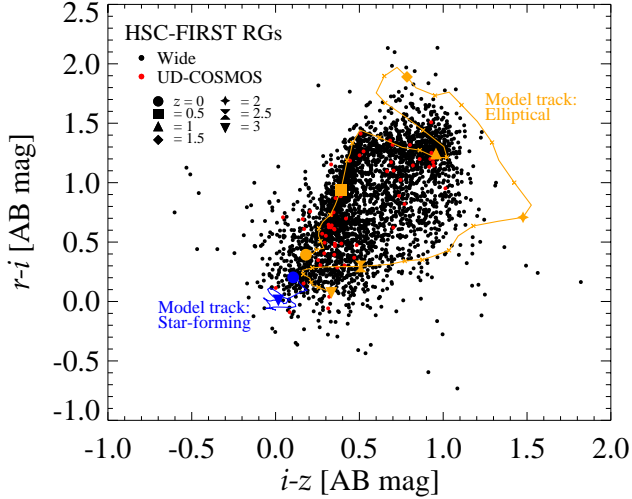}
\caption{
Color--color diagram of $r-i$ and $i-z$. 
The HSC-FIRST Wide RGs and UD-COSMOS RGs are shown as black circles and red circles.
To estimate their redshifts, color tracks of two model galaxies are overplotted:
a passive elliptical galaxy (orange curve) and a star-forming galaxy (blue curve).
Each redshift of a model track is marked by large filled symbols (see the inserted legends).
To be eye friendly, minor ticks indicated by small crosses show $\Delta z=0.1$ steps.
Only $z=0$ and $3$ marks are shown in the star-forming galaxy track.
\label{fig:riz}}
\end{figure}

The redshifts of the HSC-FIRST RG samples in both the Wide and UD-COSMOS layers are
examined in a color-color plane as the first step.
Our RG samples are shown in the $r-i$ and $i-z$ color-color diagram of Figure \ref{fig:riz}.
As the selection criteria have been imposed on the HSC-FIRST sample 
to detect in these three bands of $r$, $i$, and $z$,
the three bands' photometries are available for all of the HSC-FIRST sources.
The colors were compared with two color tracks of two model galaxies \citep{Bruzual2003}
in order to estimate their redshifts.
One is a passive elliptical galaxy that has star formation history with 
an exponential decay time of 1~Gyr, a galaxy age of 8~Gyr, and the solar metallicity.
The other is a star-forming galaxy that has an instantaneous burst, a galaxy age of 0.025~Gyr,
and the solar metallicity.
From Figure \ref{fig:riz}, we found that most of the sources are located along the track of the 
model elliptical galaxy from $z\sim 0$ to $z\sim 1$.

\subsubsection{Photometric Redshifts}\label{sec:photoz}
\begin{figure}[t!]
\epsscale{1}
\plotone{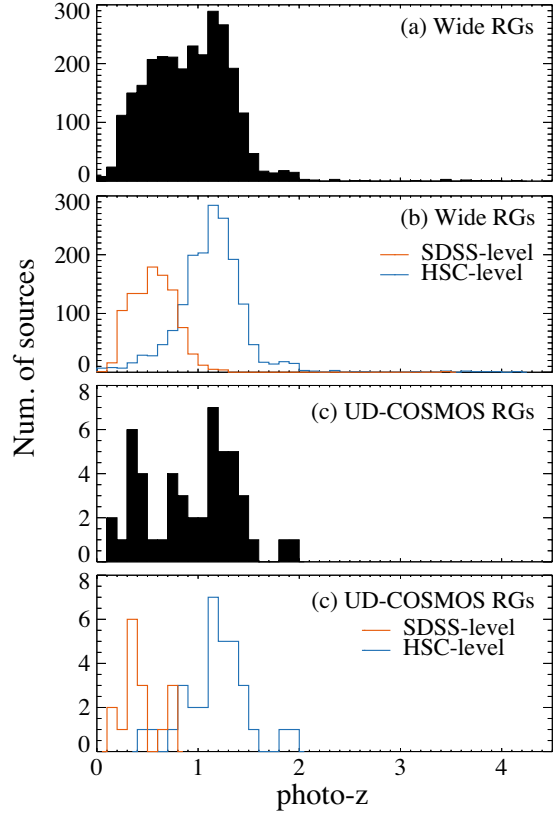}
\caption{
Histograms of photometric redshifts.
The RGs in the Wide (a) and the UD-COSMOS (c) layers are shown by black bars.
The RGs in the both layers are separated into the SDSS level (orange) and the HSC level (blue) 
in panels (b) and (d).
Only RG sources with the reliable photometric redshifts, 
with $z_{\rm photo}>0$, $2\sigma_{z_{\rm photo}} < 1$, and $\chi^2_{\nu} < 5$,
are shown.
\label{fig:photoz_hist}}
\end{figure}

Photo-$z$ which were estimated 
using the \verb|Mizuki| SED-fitting code and
the photometries of the five HSC-SSP bands (see \citealt{Tanaka2015, Tanaka2017} for details) 
are available for our HSC-FIRST RG sample.
A total of 2714 RGs in the Wide layer and 49 RGs in the UD-COSMOS layer have
reliable photo-$z$, where $z_{\rm photo}>0$, 
$2\sigma$ error in $z_{\rm photo}$ ($2\sigma_{z_{\rm photo}}$) is less than 1,
and a reduced $\chi^2$ is less than 5.
The photo-$z$ sample does not have a systematic bias against the optical magnitudes of sources at $i<24$.
On the other hand, 
$\sim 30\%$ of the faint HSC-FIRST RGs with $i>24$ are not included in the photo-$z$ sample
due to the low reliability of their photo-$z$.
This bias results in a lack of 36\% among RGs with $\log{\mathcal{R}_{\rm obs}}>5$. 
The redshifts of the photo-$z$ sample are shown in Figure \ref{fig:photoz_hist}
and range from $z_{\rm photo}\sim 0$ to $\sim 1.5$ in both the Wide and UD-COSMOS layers.
The apparent bimodal distribution with two peaks at $z_{\rm photo}\sim 0.6$ and 1.2 could be 
attributed to both a spiky photo-$z$ distribution of the parent HSC sample (see Figure 11 in \citealt{Tanaka2017})
and a rapid decrease of the photo-$z$ objects toward $z_{\rm photo} \sim 1.7$
where the five HSC bands can no longer straddle a redshifted 4000~\AA\ break. 
The interval of the spiky photo-$z$ distribution is roughly 0.1 and is smaller than 
the average uncertainties (0.2) in the photo-$z$ of the photo-$z$ sample. 
The photo-$z$ distributions of the SDSS-level and HSC-level RGs are clearly divided into
low and high redshifts, respectively.
This result means that the optically faint RGs or the HSC-level RGs ($\sim 1500$ sources)
exist at high redshifts ($\gtrsim 1$) 
rather than at the local universe as intrinsically optically faint objects.

\begin{figure}[t!]
\epsscale{1.1}
\plotone{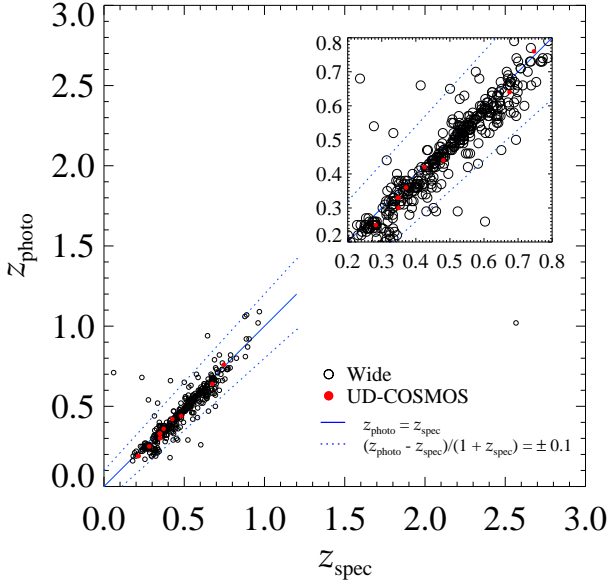}
\caption{
Comparison between the photo-$z$ and spec-$z$ of the HSC-FIRST RGs ($i < 22.0$) with reliable photo-$z$ and spec-$z$. 
The sources in the Wide (black circles) and UD-COSMOS (red circles) are shown.
The solid line represents $z_{\rm photo} = z_{\rm spec}$.
The dotted lines of $(z_{\rm photo} - z_{\rm spec})/(1+z_{\rm spec}) = \pm 0.1$ serve as a guide to the eye.
The inserted figure shows plots at a redshift range of 0.2--0.8.
\label{fig:photoz_specz}}
\end{figure}

In order to inspect the photo-$z$ for the validity, the photo-$z$ were compared with 
spectroscopically measured redshifts (spec-$z$ or $z_{\rm spec}$) when available.
We used public spec-$z$ catalogs from 
SDSS DR12 \citep{Alam2015},
GAMA DR2 \citep{Liske2015},
VIPERS PDR1 \citep{Garilli2014},
VVDS \citep{LeFevre2013},
WiggleZ DR1 \citep{Drinkwater2010},
PRIMUS DR1 \citep{Coil2011,Cool2013},
and zCOSMOS DR3 \citep{Lilly2009}.
We cross-matched the public spec-$z$ sources to the HSC-FIRST photo-$z$ sources in positions
by a $1\farcs0$ search radius. 
Of the HSC-FIRST photo-$z$ sample, 404 RGs in the Wide layer and 
11 RGs in the UD-COSMOS have secure spec-$z$
where $z_{\rm spec} > 0$, an error in $z_{\rm spec}$ is greater than 0, and 
a secure homogenized quality flag of $z_{\rm spec}$ determinations (see \citealt{Tanaka2017} for details).
The optical magnitudes of the spec-$z$ sample are limited to be $i < 21.8$ 
due to the magnitude limit of the spectroscopic observations.
In Figure \ref{fig:photoz_specz}, the photo-$z$ and the spec-$z$ are compared each other.
The means and standard deviations of the relative differences between both are
$(z_{\rm photo} - z_{\rm spec})/(1+z_{\rm spec}) = -0.015 \pm 0.057$ for the Wide sources
and $-0.017 \pm 0.012$ for the UD-COSMOS sources.
In the overall trend, the photo-$z$ is in good agreement with spec-$z$ for the $i < 21.8$ sources.

\begin{figure}[t!]
\epsscale{1.1}
\plotone{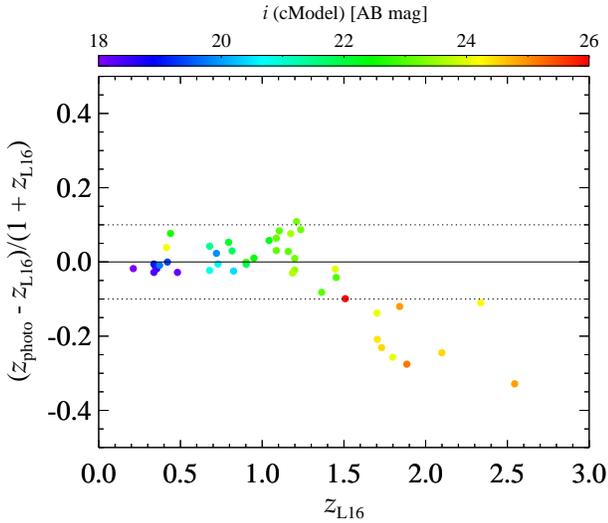}
\caption{
Relative difference between the HSC-SSP photo-$z$ ($z_{\rm photo}$)
and the photo-$z$ in the COSMOS2015 catalog ($z_{\rm L16}$, \citealt{Laigle2016}).
The plotted data are the photo-$z$ sources in the UD-COSMOS, which have 
secure photo-$z$ in both catalogs (see text).
The colors of the data indicate the $i$-band magnitude of the sources 
according to the color bar at the top of the figure.
The solid line represents $z_{\rm photo} = z_{\rm L16}$ and
the dotted lines $(z_{\rm photo} - z_{\rm L16})/(1+z_{\rm L16}) = \pm 0.1$ as a guide to the eye.
\label{fig:photoz_cosmos}
}
\end{figure}

In order to examine the photo-$z$ of the faint sources with $i > 22.0$, 
we utilized the photo-$z$ ($z_{\rm L16}$) of the COSMOS2015 catalog in the COSMOS field, 
which were estimated using thirty multiband photometric data from the ultraviolet to mid-infrared \citep{Laigle2016}.
The UD-COSMOS photo-$z$ RG subsample includes 44 counterparts of the COSMOS2015 sources with
the minimum reduced $\chi^2$ of the SED fitting being less than 5. 
In Figure \ref{fig:photoz_cosmos},
the relative differences between $z_{\rm photo}$ and $z_{\rm L16}$ for 44 matched RGs are shown as
a function of $z_{\rm L16}$. 
Both the photo-$z$ are consistent with each other within 10\% up to $z_{\rm L16}\sim 1.7$.
The standard deviation of the relative difference is 0.047 in $z_{\rm L16} < 1.7$.
At $z_{\rm L16}>1.7$, where the sources are fainter than $i\sim 24$,
the relative difference is systematically negative
and is $-30\%$ at the most. 
This deviation at $z_{\rm L16}>1.7$ is attributed to a redshifted 4000~\AA\ break beyond the $y$-band,
and therefore, the photo-$z$ of these sources cannot be accurately constrained only from 
the photometries of the five HSC-SSP bands.
These examinations show that the HSC-SSP photo-$z$ could be secure, 
at least, at $z_{\rm photo} \lesssim 1.7$, 
although the photometric sensitivities of the UD-COSMOS layer are deeper than those the Wide layer.

\subsubsection{Optical and Radio Properties}\label{sec:discussion_properties}
\begin{figure}[t!]
\epsscale{0.9}
\plotone{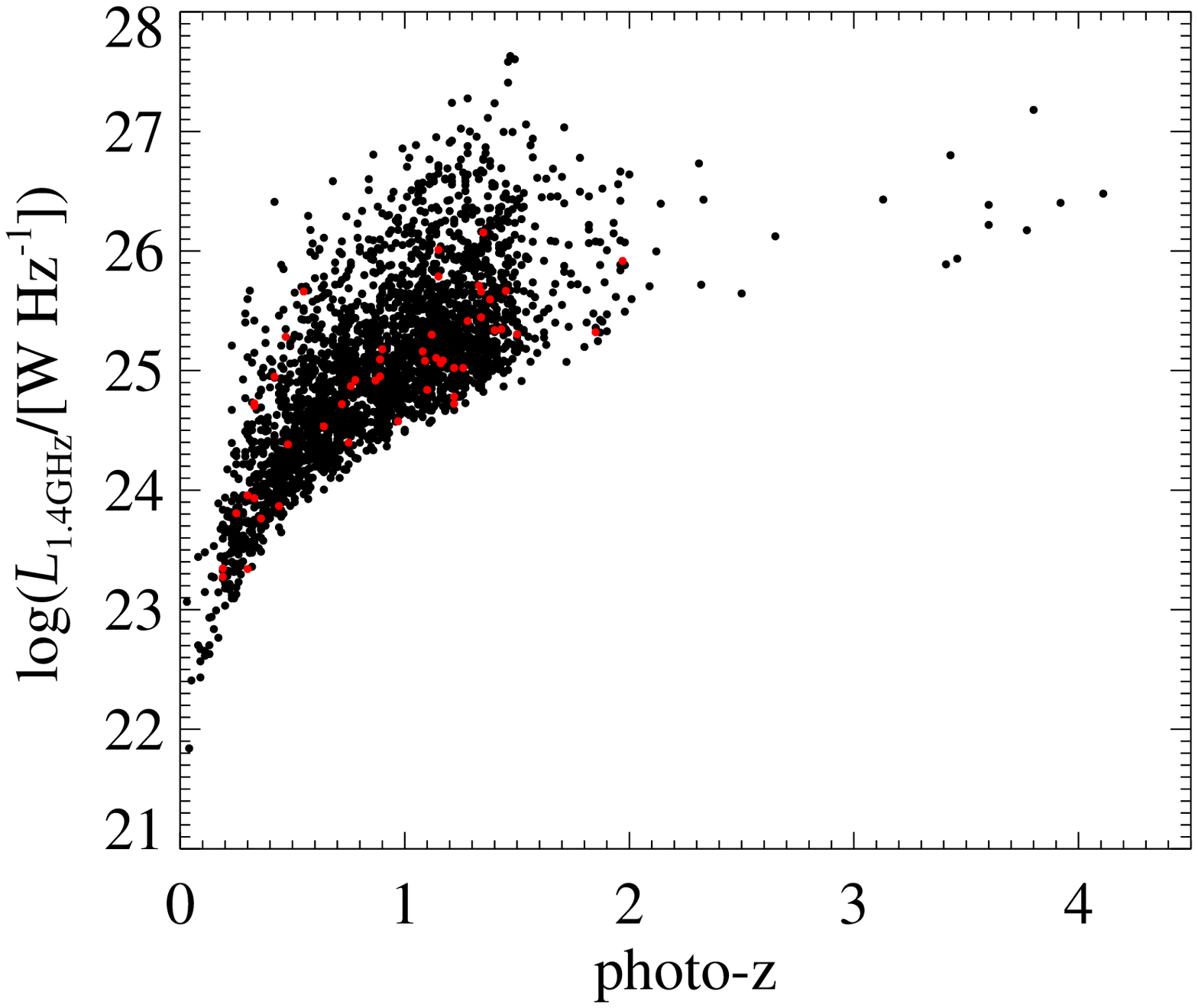}
\plotone{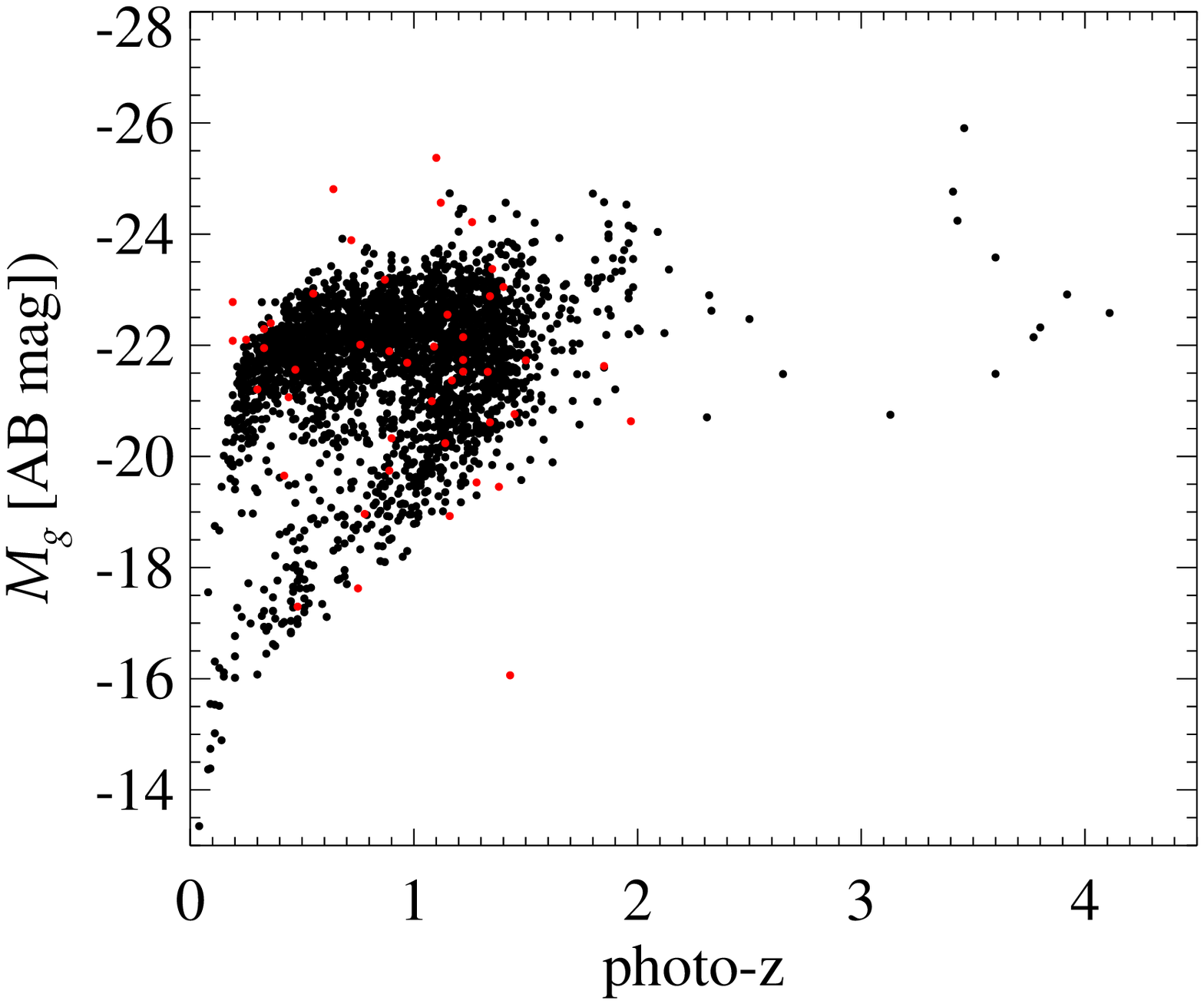}
\caption{
The rest-frame 1.4GHz radio luminosity (upper panel) 
and the absolute $g$-band magnitude at the rest frame (lower panel) 
as a function of photo-$z$.
Only RG sources with the reliable photometric redshifts, 
with $z_{\rm photo}>0$, $2\sigma_{z_{\rm photo}} < 1$, and $\chi^2_{\nu} < 5$,
are shown.
The black circles and red circles denote the Wide and UD-COSMOS RGs, respectively.
\label{fig:photoz_Mag}}
\vspace{5mm}
\end{figure}

The optical luminosity and radio luminosity at the rest frame
were calculated using the photo-$z$ for the photo-$z$ subsample.
In Figure \ref{fig:photoz_Mag}, the calculated luminosities are shown as a function of photo-$z$.
The 1.4~GHz radio luminosity was 
$k$-corrected by assuming a power-law radio spectrum of the form $\propto{\nu^{\alpha}}$ with 
index $\alpha=-0.7$ (e.g., \citealt{Condon1992}).
The sources at $z\gtrsim 1$ have high 1.4~GHz luminosities compared with lower redshift sources
due to a bias from the FIRST sensitivity.
The rest-frame $g$-band absolute magnitude was estimated by the SED template fitting \verb|Mizuki|
\citep{Tanaka2015, Tanaka2017}.
There may be two sequences of the $g$ magnitude along the redshifts. 
The bright sequence with $M_g \lesssim -21$ does not strongly depend on the redshift.
On the other hand, the number of sources belonging in the faint sequence seems to dramatically increase with increasing redshifts.
The sources with faint absolute magnitudes are also faint in apparent magnitudes ($i>24$),
and the optical SEDs of the HSC-SSP bands show that some sources on the faint sequence have 
a flat SED without distinct features. 
Thus, the redshifts of these sources are probably estimated incorrectly. 
In fact, the comparison of $M_g$ and redshifts with the COSMOS2015 catalog \citep{Laigle2016}
in UD-COSMOS shows the underestimated redshifts of the faint sequence sources, 
and the redshifts of the COSMOS2015 indicate that most sources in the faint sequence 
should be at $z_{\rm L16} > 1.5$.
In this comparison, a source at $z = 0.5$ on the faint sequence was found
to have redshift and $M_g$ consistent with the COSMOS2015 catalog. 
This source is really a local source with a faint optical luminosity.
The faint sequence could be an apparent one due to
both the effects of the underestimated redshifts and 
the Malmquist bias of the HSC-SSP sensitivity limits.
Some sources at $z \lesssim 0.5$ may be local galaxies with strong radio radiation 
from active star formation.

\begin{figure}[t!]
\epsscale{1.1}
\plotone{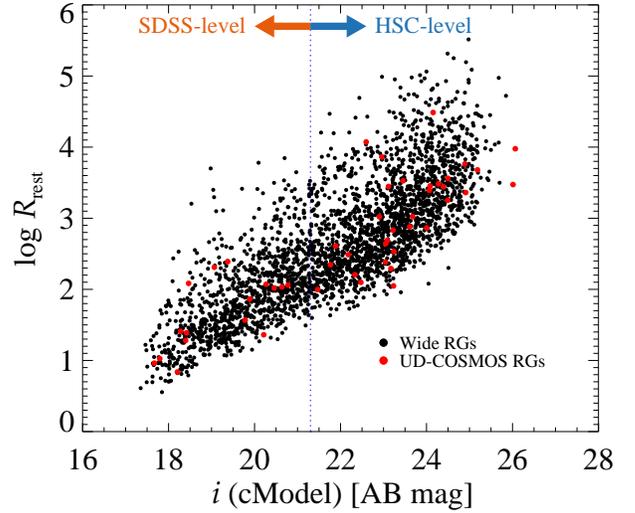}
\caption{
Radio loudness at the rest frame as a function of the apparent $i$-band magnitude.
The RG sources with reliable photo-$z$ in the Wide (black) and the UD-COSMOS (red) layers are shown.
The blue dotted line is the border separating the SDSS-level and HSC-level samples 
by their $i$-band magnitude.
\label{fig:imag_rest_radioloudness}}
\end{figure}

Radio loudness at the rest frame was estimated and is defined here as 
a ratio of the rest-frame 1.4~GHz flux to the rest-frame $g$-band flux.
The wavelength of the rest-frame $g$-band flux is close to the rest-frame
$B$-band, which is typically used for calculating the radio loudness.
The rest-frame radio loudnesses of the SDSS- and HSC-level sources as a function of
the apparent $i$-band magnitude are shown in Figure \ref{fig:imag_rest_radioloudness}.
We found that the RGs with 
high radio loudness ($\log{R_{\rm rest}} > 3$) are found in the HSC-level subsample.
Certainly, there are likely also low radio-loudness RGs in the HSC-level regime,
but such sources are not included due to the sensitivity limit of the FIRST survey.
Note that the radio loudness we use here is underestimated, 
because HSC band flux could be dominated by stellar light from host galaxies of RGs,
and the rest-frame $g$-band flux is derived from the SED templates of galaxies.

\begin{figure}[t!]
\epsscale{1.1}
\plotone{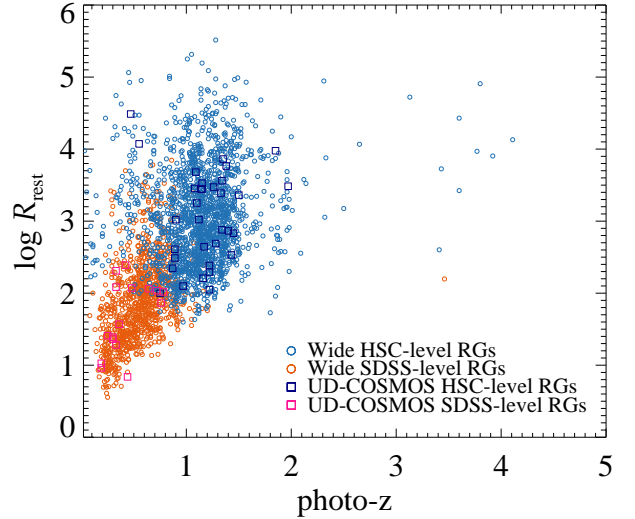}
\caption{
Radio loudness at the rest frame as a function of photo-$z$.
Only the RG sources with reliable photo-$z$ are shown.
Sources at the SDSS-level and the HSC-level brightness in the Wide layer (UD-COSMOS)
are represented by blue circles (dark blue squares) and orange circles (pink squares), respectively.
\label{fig:photoz_Rrest}}
\end{figure}

We show each distribution of the SDSS- and HSC-level RG samples 
in an $R_{\rm rest}-z_{\rm photo}$ plane in Figure \ref{fig:photoz_Rrest}.
We distinctly found that
the optically faint RGs (the HSC level) have higher radio loudnesses
and photo-$z$ than the optically bright RGs (the SDSS level). 
Again, we note that optically faint RGs with a low radio loudness 
are not detected due to the FIRST sensitivity. 
However, we claim that
our HSC-FIRST RG sample includes a large number ($\sim 730$) of 
the optically faint RGs that have high radio loudness ($\log{R_{\rm rest}} > 3$)
and redshifts of more than 1.
These RGs have not been probed by SDSS.

\subsection{Optical Spectra and Quasar/Galaxy Classification}\label{sec:spectra}
We performed a quasar/galaxy classification using optical spectra
in order to inspect the morphological classification based on the HSC-SSP images 
in Section \ref{sec:optical_morphology}.
This inspection is based on whether or not a spectrum of a source that was morphologically 
classified as a quasar (or a galaxy) shows broad emission lines.
We used archival spectra available from SDSS DR12 \citep{Alam2015}, 
GAMA PDR2 \citep{Liske2015}, and zCOSMOS \citep{Lilly2009}.
In addition to the archival spectra, 
we also examined 18 spectra that have been recently obtained 
with 2dF-AAOmega installed on 3.9~m Anglo-Australian Telescope (AAT) in 2017 July. 
See He et al.\ (2018, in preparation) for the details about
the observation and data reduction.
Finally, we inspected 622 sources in the Wide layer and 21 in the UD-COSMOS layer.
These subsamples include sources without a secure spec-$z$. 
Although this spectral inspection allows us to confirm 
the morphological classification using the HSC-SSP images,
we should take into account the following two points.
One is that the $i$-band magnitudes of these subsamples are biased toward
a bright side and are distributed from 17.4 to 22.8 in the Wide layer and 
from 17.7 to 21.9 in UD-COSMOS. 
The other is that this spectral inspection does not distinguish between 
quasars and broad-line RGs.

\begin{figure*}[t!]
\epsscale{1.15}
\plotone{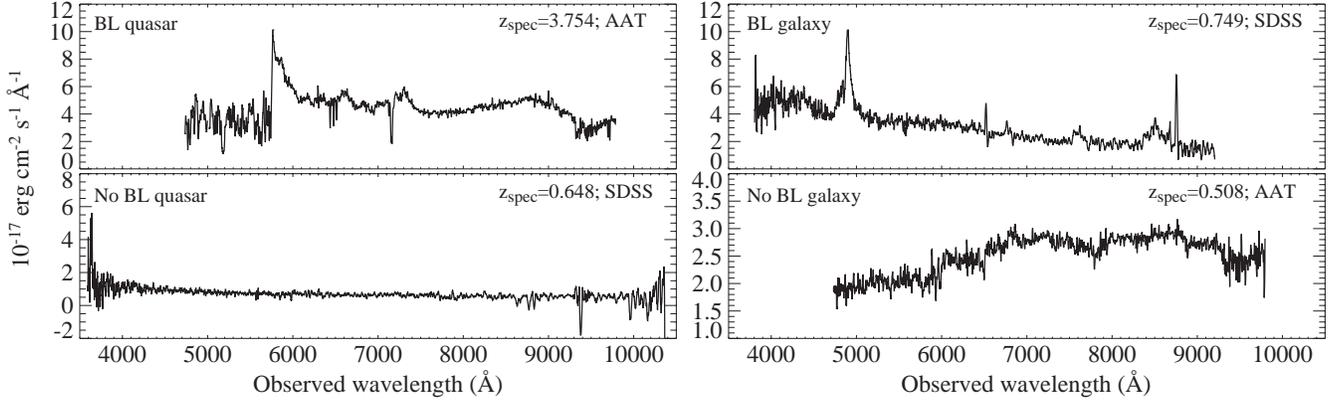}
\caption{
Examples of optical spectra of HSC-FIRST sources.
Four types of spectra are shown: 
a morphologically classified quasar with broad emission lines (top left),
a morphologically classified galaxy with broad emission lines (top right),
a morphologically classified quasar without broad emission lines (bottom left), and
a morphologically classified galaxy without broad emission lines (bottom right).
Two spectra were obtained with AAT (He et al.\ 2018, in preparation) and
others were taken from SDSS DR12.
\label{fig:spectra}}
\end{figure*}

Of the 622 Wide sources with available spectral classification,
we have 152 morphologically classified quasars and
470 galaxies. 
The inspection results show that 
130 (86\%) morphologically classified quasars 
and 26 (5.5\%) morphologically classified galaxies show broad emission lines.
In UD-COSMOS, 
2/2 (100\%) morphologically classified quasars
and 1/19 (5.3\%) morphologically classified galaxies
show broad emission lines.
Therefore, 
at $i\lesssim23$, 
there could be 5\% contaminations of quasars in both the morphologically classified RG samples
of the Wide layer and UD-COSMOS. 
Examples of four types of spectra, 
a quasar with/without broad emission lines 
and a galaxy with/without broad emission lines,
are shown in Figure \ref{fig:spectra}.

\begin{figure}
\vspace{5mm}
\epsscale{1}
\plotone{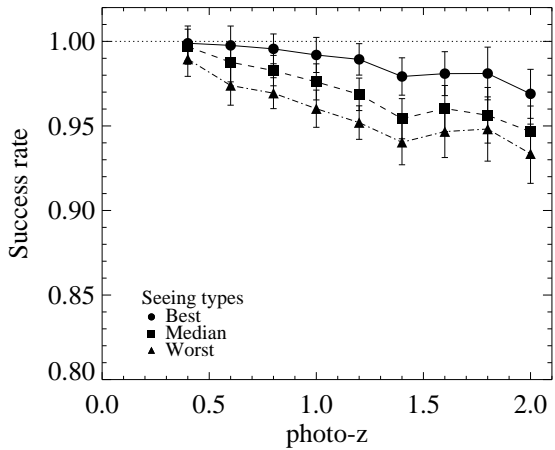}
\caption{
Success rate of the HSC morphological classification of galaxy/quasar as a function of redshift. 
The success rate is defined as 
the fraction of a number of sources classified as a galaxy by both HST/ACS and HSC to
the number of sources classified as a galaxy by HST/ACS.
Three types of seeing in the HSC images are shown:
best (FWHM $\sim 0\farcs5$, circle,), median ($\sim 0\farcs7$, square), and worst ($\sim 1\farcs0$, triangle). 
The error bars represent the Poisson errors.
\label{fig:hstacs}}
\end{figure}

Additionally, we inspected a redshift dependence of the success rate
of the morphological classification with the HSC images. 
In the classification, an RG could be misclassified as a radio-loud quasar 
when a source is at a high redshift. 
We used the morphological information measured with the Advanced Camera for Surveys (ACS)
on the {\it Hubble Space Telescope} (HST) in the COSMOS region \citep{Leauthaud2007}
and our morphological classification with simulated HSC images at the Wide layer depth in UD-COSMOS.
This inspection employed HST ACS morphology data and HSC UD-COSMOS 
at an imaging depth equivalent to the Wide-layer depth.
The ACS images have a higher spatial resolution ($0\farcs12$) than the HSC Wide layer
and thus provide a robust classification.
The HSC UD-COSMOS images at the Wide layer depth provide three types of seeing: 
best, median, and worst seeing images with FWHMs of $0\farcs5$, $0\farcs7$, and $1\farcs0$, respectively \citep{Aihara2017b}.
The success rate of the HSC morphological classification is defined as 
a fraction of a number of sources classified as a galaxy by both ACS and HSC to
a number of sources classified as a galaxy by ACS.
The success rate was examined up to $z_{\rm photo} = 2$ as presented in Figure \ref{fig:hstacs}.
We confirmed the redshift dependence of the success rate, 
which moderately decreases with redshifts.
The success rate is 95--97\% at $z_{\rm photo} > 1$ in the median seeing. 
Therefore, we found that 3--5\% of RGs at $z_{\rm photo}>1$ could be radio-loud quasars.

\subsection{Improved matching rate}\label{sec:high_matching_rate}
Our study using the Subaru HSC-SSP catalog has successfully identified
the optical counterparts of 51\% and 59\% of the FIRST sources in the HSC-SSP Wide layer
and UD-COSMOS, respectively.
These matching rates are significantly higher than the $\sim 30\%$ matching rates of
the SDSS-based studies by \citet{Ivezic2002}, \citet{Helfand2015}, and \citet{Ching2017}.
Although the adopted search radii are different between the studies, 
even if we use the completeness-corrected matching rate, 
our result (47\% in the Wide layer) is still higher than the corrected rate of 31\% in \citet{Ivezic2002}.
It is clear that the improved matching rate is due to the utilization of the deep optical catalog,
compared with the previous studies, 
from Figure 
\ref{fig:i_numbercounts}, \ref{fig:radioloudness}, and \ref{fig:imag_rest_radioloudness}.
This interpretation is supported by the higher matching rate of 59\% in UD-COSMOS,
which has fainter HSC-SSP sources than the Wide layer.

However, approximately 40\% of FIRST sources in the HSC-SSP unmasked region still have 
no optical counterparts within 1\arcsec\ of their positions.
Our sample criterion of ${\rm S/N}>5$ in $r$, $i$, and $z$ likely 
misses a certain fraction of HSC-SSP sources with an extremely red color at the faint end,
such as high-$z$ red galaxies.
Other unmatched FIRST sources are roughly grouped into two types
from the inspections of their HSC-SSP images.
The first consists of radio sources
with bright radio companion(s) close to it.
These sources are likely radio lobes associated with low-$z$ RGs.
The second consists of single radio components.
These optical counterparts may be extremely faint in the optical ($i<26$),
such as radio AGNs with a quite low accretion rate, 
obscured AGNs, radio AGNs in less massive host galaxies, or high-redshifted objects.


\section{Summary}\label{sec:summary}

We presented the results of the exploration of radio AGNs using 
the Subaru HSC-SSP catalog and the FIRST VLA catalog.
This is the initial result of an ongoing project, WERGS.
The positional cross-match between HSC-SSP and FIRST produced
3579 and 63 optical counterparts of the FIRST radio sources
in the Wide layer and UD-COSMOS, respectively.
These numbers of the matched counterparts correspond to more than 50\% 
of the FIRST radio sources in the search fields, showing 
a much higher matching rate than the previous studies of optically bright
SDSS-FIRST sources ($i<21$).

We reported the properties in the optical and radio wavelengths of 
the matched HSC-FIRST sources.
Among the HSC-FIRST sources, 9\% in the Wide and 3\% in the UD-COSMOS layers
are optically unresolved sources, that is, likely radio-loud quasars.
The $i$-band number counts show a flat slope down to 24~mag, implying
a constant number of either less massive or distant radio AGNs. 
The optically faint sources show the different slopes of the 1.4~GHz 
source counts from the bright sources with $i<21.3$.
Almost all of the photometric redshifts of the subsample with available redshift data
range up to 1.5. In particular, the optically faint sources are distributed 
around the redshift of 1.
The comparison of the color-color diagram with the galaxy model also shows
a similar redshift distribution to the photometric one, 
and most of the sources are consistent with redshifted early-type galaxies 
although there is a large dispersion.
Thanks to the deep and wide HSC-SSP survey, 
we found a large number ($\sim 730$) of sources with high radio loudness ($\log{R} > 3$) 
in the optically faint regime. Such sources are distributed at $z\gtrsim 1$.

We also showed that the rare objects such as sources with the highest radio loudness 
were found in the wide and shallow field (the Wide layer) rather than in the small and 
deep field (UD-COSMOS). 
Together with the improved matching rates,
these results demonstrated the advantages of our project over the previous radio AGN searches 
in terms of both optical depth ($i<26$) and 
the wide search field (more than 150~deg$^2$).

\acknowledgments

We would like to thank the anonymous referee for very
useful comments that helped to further improve this paper.

This work is based on data collected at the Subaru Telescope and retrieved from the HSC data archive system, which is operated by the Subaru Telescope and Astronomy Data Center at the National Astronomical Observatory of Japan.

The Hyper Suprime-Cam (HSC) collaboration includes the astronomical communities of Japan and Taiwan, and Princeton University.  The HSC instrumentation and software were developed by the National Astronomical Observatory of Japan (NAOJ), the Kavli Institute for the Physics and Mathematics of the Universe (Kavli IPMU), the University of Tokyo, the High Energy Accelerator Research Organization (KEK), the Academia Sinica Institute for Astronomy and Astrophysics in Taiwan (ASIAA), and Princeton University.  Funding was contributed by the FIRST program from the Japanese Cabinet Office, the Ministry of Education, Culture, Sports, Science and Technology (MEXT), the Japan Society for the Promotion of Science (JSPS), Japan Science and Technology Agency  (JST),  the Toray Science  Foundation, NAOJ, Kavli IPMU, KEK, ASIAA,  and Princeton University.

This paper makes use of software developed for the Large Synoptic Survey Telescope. 
We thank the LSST Project for making their code available as free software at http://dm.lsstcorp.org

The Pan-STARRS1 Surveys (PS1) have been made possible through contributions of the Institute for Astronomy, the University of Hawaii, the Pan-STARRS Project Office, the Max Planck Society and its participating institutes, the Max Planck Institute for Astronomy, Heidelberg, and the Max Planck Institute for Extraterrestrial Physics, Garching, The Johns Hopkins University, Durham University, the University of Edinburgh, Queen's University Belfast, the Harvard-Smithsonian Center for Astrophysics, the Las Cumbres Observatory Global Telescope Network Incorporated, the National Central University of Taiwan, the Space Telescope Science Institute, the National Aeronautics and Space Administration under grant No. NNX08AR22G issued through the Planetary Science Division of the NASA Science Mission Directorate, the National Science Foundation under grant No. AST-1238877, the University of Maryland, and Eotvos Lorand University (ELTE)

Based in part on data acquired through the Australian Astronomical Observatory, 
under program A/2017A/008. 
We acknowledge the traditional owners of the land on which the AAT stands, 
the Gamilaraay people, and pay our respects to elders past and present.

The National Radio Astronomy Observatory is a facility of the
National Science Foundation operated under cooperative agreement
by Associated Universities, Inc.

This work is financially supported by the Japan Society for the
Promotion of Science (JSPS) KAKENHI 16H01101, 16H03958, and 17H01114 (TN).
Y.M. was supported by the Japan Society for the Promotion of Science (JSPS)
KAKENHI Grant No. JP17H04830 and the Mitsubishi Foundation Grant No. 30140.

\vspace{5mm}
\facilities{Subaru (HSC), VLA.}

\software{STILTS \citep{Taylor2006}, IDL, IDL Astronomy User's Library \citep{Landsman1993}, TOPCAT \citep{Taylor2005}.
          }

\appendix
\section*{Counting FIRST Sources on the Unmasked HSC-SSP Regions}\label{app:count}

In order to count the number of FIRST sources on the unmasked HSC-SSP regions,
we have to consider to the FIRST sources lying within the masked HSC-SSP regions
that result from the selection of the clean HSC-SSP sample.
The flags in the masked regions are described in mask images through an integer bitmask value
(see \citealt{Bosch2017} for details).
Regions with the following bitmasks are considered as masked regions
corresponding to the sample selection criteria described in Section \ref{sec:HSCdata}:
\verb|BAD| (a sensor defect), 
\verb|BRIGHT_OBJECT| (a region near a very bright object), 
\verb|CR| (a cosmic ray), 
\verb|EDGE| (the edge of a CCD or coadded patch),
\verb|INTRP| (an interpolated pixel),
\verb|SAT| (a saturated pixel),
and \verb|SUSPECT| (a pixel value above the level where the linearity correction is reliable).

There were originally 8282 (118) FIRST sources in the HSC-SSP Wide layer (UD-COSMOS) 
defined in Table \ref{tab:HSCclean} before any masks were performed.
These numbers were FIRST sources which have the nearest HSC-SSP sources within 20\arcsec
for the purpose of excluding FIRST sources on the patches with poor photometries 
(see Section \ref{sec:HSCdata}) and counting FIRST sources in UD-COSMOS.
After masking, we obtained 7072 and 106 FIRST sources in the unmasked regions of 
the Wide layer and UD-COSMOS, respectively.

\end{document}